\documentclass[final,3p,times]{elsarticle}
\makeatletter
\def\ps@pprintTitle{%
 \let\@oddhead\@empty
 \let\@evenhead\@empty
 \def\@oddfoot{Forthcoming in Transportation Research Part C\hfill}%
 \let\@evenfoot\@oddfoot}
\makeatother
\usepackage[hidelinks]{hyperref}
\usepackage{siunitx}

\usepackage{xcolor}
\usepackage{geometry}
\usepackage[colorinlistoftodos]{todonotes}
\usepackage{float} 
\usepackage{caption}
\usepackage{setspace,etoolbox,threeparttable,array,subcaption,colortbl,makecell,multirow,soul,color,hhline,rotating,hanging,ragged2e,diagbox,booktabs,multicol}
\usepackage[pagewise]{lineno}

\AtBeginEnvironment{tablenotes}{\singlespacing}

\newcommand{\RoundTwo}[1]{\textcolor{black}{#1}}

\newcommand{\smallp}{\RoundTwo{$<$0.01}}
\newcommand{\sig}{\color{red}*\color{black}}

\begin{document}

\begin{frontmatter}

\title{\textbf{Fundamental Diagrams of Commercial Adaptive Cruise Control: Worldwide Experimental Evidence}} 


\author[add1]{Tienan Li} 
\author[add1]{Danjue Chen\corref{cor1}} 
\author[add2]{Hao Zhou}
\author[add1]{Yuanchang Xie} 
\author[add2]{Jorge Laval}
\address[add1]{Civil and Environmental Engineering, University of Massachusetts, Lowell}
\address[add2]{Civil and Environmental Engineering, Georgia Institute of Technology}
\cortext[cor1]{Corresponding author: $danjue\_chen@uml.edu$}

\begin{abstract}

Experimental measurements  on commercial adaptive cruise control (ACC) vehicles \RoundTwo{are}  becoming increasingly available from around the world,  providing an unprecedented opportunity to study the traffic flow characteristics that arise from this technology.
This paper adds new experimental evidence to this knowledge base and presents a comprehensive empirical study on the ACC equilibrium behaviors via the resulting fundamental diagrams. We find that like human-driven vehicles, ACC systems display a linear \RoundTwo{equilibrium} spacing-speed relationship (within the range of available data) but the key parameters of these relationships can differ significantly from human-driven traffic depending on input settings: At the minimum headway setting, \RoundTwo{equilibrium} capacities in excess of 3500 vehicles per hour are observed, together with an extremely fast \RoundTwo{equilibrium wave speed of 100 kilometers per hour on average}. 
These fast waves are unfamiliar to human drivers, and \RoundTwo{may} pose a safety risk. \RoundTwo{The results also suggest} that ACC jam spacing \RoundTwo{can be} much larger than in human traffic,  which reduces the network storage capacity\RoundTwo{}. 


\end{abstract}
\end{frontmatter}






\section{\textbf{Introduction}}

Adaptive Cruise Control (ACC) technology, as a precursor of automated vehicle technology, has been available for over two decades and is now adopted in the car models from more than 20 manufacturers \citep{wikipedia_2020}. According to \cite{kyriakidis2015deployment}, about 5\% of newly sold cars are equipped with ACC and the proportion is steadily growing. A comprehensive review of the market ACC systems and their service providers can be seen in \cite{zhou2021review}. It can be expected that the penetration of ACC vehicles will continue to grow and become very significant, and thus will have profound impacts on traffic flow.  The literature has claimed that ACC has the potential to improve traffic flow such as increasing throughput and traffic stability  \citep{van2006impact,kesting2008adaptive,shladover2012impacts,delis2015macroscopic,talebpour2016influence,zheng2020analyzing,huang2020scalable,sun2020relationship}. However, most of those claims are based on simulation outcomes and lack the empirical ground. 

Understanding the equilibrium behavior of ACC is fundamental to understanding their impacts on the congestion characteristics of urban networks of the future. But existing literature focuses on the string instability of   ACC systems to develop new controllers \citep{naus2010string, ploeg2011design, bu2010design, milanes2013cooperative, milanes2014modeling}.  For example, \cite{naus2010string} tested Citroen C4s and found that it had worse string stability compared to the cooperative adaptive cruise control (CACC) newly proposed. 
\cite{bu2010design} found that the ACC on Infinity FX45s had large time gap variation, which implied potential loss of string stability. 
Since the studies mentioned above mainly aimed to compare the string stability of commercial ACC and the new controllers, the experiments often tested only a limited set of the traffic conditions (like speed), such as the string stability under one headway setting in a small speed range.  

Fortunately, some recent empirical tests have emerged to investigate the ACC behaviors in a larger variety of traffic conditions, including non-equilibrium and equilibrium conditions.  Specifically, a recent experiment conducted by \citep{victor2019platoon} tested a seven-ACC platoon on public roads for a 500-km trip. It was observed that the ACC platoon can be cut in by surrounding human-driven vehicles even with the smallest headway setting, implying that the ACC vehicles may have a relative large time gap. 
\cite{shi2021empirical} recently conducted a set of three-vehicle platoon experiments with two commercial Lincoln MKZ ACC cars. It was found the platoon became more unstable as the headway was set to a smaller value. The collected data were fitted to a simple linear controller \citep{li2020tradeoff}
. The model estimation results for key parameters were found not consistent over different speed ranges, which suggested the potential nonlinearity and stochasticity in the ACC controller.

Another important pilot experiment was conducted by Dr. D. Work's research team and their collaborators, which involved a few field tests on eight different ACC car models  \citep{gunter2019model,gunter2020commercially}, referred to as the \textit{Work Experiments}.  With a primary focus on string stability, the \textit{Work Experiments} 
tested vehicles in a set of car-following (CF) scenarios involving different headway settings, magnitude of disturbances and driving speeds. The data were then used to calibrate the ACC controller for each car model via a modified optimal velocity model \citep{bando1995dynamical}.  Simulations based on the calibrated models indicate that all eight ACC  systems are string unstable. 
Another important milestone is the ACC campaign of the Joint Research Centre (JRC) of European Commission \citep{makridis2021openacc}, referred to as the \textit{OpenACC Experiments}, which has conducted a set of experiments to reveal the impacts of ACC on traffic flow and energy consumption.  More than 20 ACC car models were tested. Notably, the JRC tests had varying platoon settings (from two-car to 10-car platoons), different speed levels, and different perturbations, but all the ACC systems tested were found unstable in all the perturbation events \citep{he2020energy, makridis2020empirical}. It was also found that the ACC response time varied from \SI{0.6}{s} to \SI{3.7}{s}, 
 comparable to the values of human drivers, implying comparable traffic flow if the traffic flow consists of ACC vehicles.  
The serial JRC efforts have revealed important features of ACC and provided valuable data to study ACC behaviors.  However, in-depth analysis of the ACC behaviors based on this valuable dataset is not yet available.  
Recently, we have conducted extensive field experiments using four different commercial ACC car models (referred to as the \textit{MA Experiments}) and investigated their CF behaviors under disturbances \cite{li2021ACC}. It was found that ACC behaviors were complex and they largely depend on headway settings, speed levels, and the features of stimulus from the lead vehicle.  However, the behaviors of ACC in steady traffic were not fully studied yet.  

It is worth noting that, some naturalistic driving studies involving commercial ACC vehicles have been conducted too. For example, \cite{alkim2007field} and \cite{viti2008driving} tested 20 Volkswagen Passats in public traffic for different speed levels. The results found the speeds and headway variance were lower when ACC was used (in comparison to the time when ACC was not used), but the average headway increased. \cite{schakel2017driving} used four different ACC car models and found that ACC vehicles increased spacing and time headway in saturated conditions (60-\SI{100}{km/h}) compared to human driving, suggesting a potential reduction in capacity. The research along this line reveals important behavior changes caused by ACC and potential impacts on traffic flow.  Note that, these studies mostly conducted the analysis in an aggregate manner.  For example, in \cite{schakel2017driving}, the spacing difference (with ACC on and off) was estimated across all drivers in a large speed range (60-\SI{100}{km/h}), and \cite{alkim2007field} only analyzed the headway features under three discrete traffic levels, free ($>$\SI{90}{km/h}), congested ($<$\SI{40}{km/h}), and busy (in between). Therefore, it is still not yet clear how ACC will behave in different traffic conditions and how that impact will traffic flow.

Unlike the ACC systems, human-driven vehicles (HDVs) are well studied. It is well recognized that the behaviors of HDVs have large variation.  For example, the classic Kinematic Wave model \citep{lighthill1955kinematic,richards1956shock}, which uses a triangular shape fundamental diagram (FD), can capture the basic traffic flow features in equilibrium.  The simplified Newell's car-following model \citep{newell2002simplified} is consistent with such a triangular shape fundamental diagram.  The literature has shown that the Kinematic Wave model and Newell's simplified model actually well capture the average behavior of HDVs \citep{ahn2004verification}. 
A lot of modifications have been made to better represent the HDVs. For example, a reverse-$\lambda$ shape FD \citep{koshi1983some} has been generally observed in  empirical data and associated with capacity drop. 
Furthermore, the literature has examined different factors that contributes to the variation in FD. For example,  \cite{zhang2002kinematic} extended the Kinematic Wave model to describe traffic with multi-class vehicles. 
\cite{coifman2015empirical} and \cite{ponnu2017adjacent} investigated the impact of vehicle length and inter-lane dependency on FDs, respectively.  \cite{wang2013stochastic} proposed a stochastic model for the equilibrium traffic. \cite{qu2017stochastic} proposed an alternated stochastic model, which was applicable to generate explicit stochastic speed-flow relationships with scattered samples.  A more detailed review of the HDV traffic in equilibrium can be seen in \cite{wang2013stochastic}. 

Clearly, while the literature has provided some valuable results on the equilibrium behaviors of commercial ACC systems, in-depth analysis is still missing.  Note that among the studies that involved the equilibrium behaviors of ACC vehicles, a very common practice is to calibrate a controller for the ACC and then extracts the equilibrium component; i.e., the equilibrium behavior of ACC is extracted \textit{indirectly}.  This is probably inspired by the conventional design of controllers, where a controller usually consists of two elements, a spacing policy that describes the equilibrium condition and an acceleration function that describes how ACC reacts once it deviates from the equilibrium.  Thus, the spacing policy extracted from the calibrated controller yields the ACC equilibrium behavior.  For example, this was practiced in \cite{gunter2020commercially,gunter2019model}.  
Such a practice seems reasonable, but there are some limitations.  Firstly, in reality, the controller design is often proprietary and the controller structure is unknown. Secondly, vehicle behaviors may not be exactly the same with the controller design due to the errors in low-level execution \citep{zhou2021impact}. Thirdly, for controllers designed from a non-conventional perspective, such as Artificial Intelligent (AI) based, there may not be an explicit equilibrium formulation embedded. For example, \cite{morton2016analysis} used a recurrent neural network to train the car-following (CF) behaviors.
\cite{zhu2020safe, qu2020jointly, jiang2021dampen} used reinforcement learning to generate CF behaviors with different rewards. \cite{bojarski2016end} applied an end-to-end approach that trains the vehicles from videos. Note that the AI-based approach is becoming more and more popular for Level-3 and above automated vehicles \citep{bojarski2016end,bansal2018chauffeurnet_waymo}. 
Considering these potential uncertainties in controller design and execution, in this paper, we study the equilibrium behaviors \textit{directly}; i.e., we directly measure the speed-spacing relationships when the vehicles are in equilibrium conditions.     

In this paper, we aim to understand the equilibrium behaviors of ACC vehicles based on empirical experiments and predict the impacts of ACC on traffic flow (e.g., capacity, jam density and wave speed).  This is realized by analyzing all the available experimental evidences worldwide, including our recent experiments\citep{li2021ACC} and the rich datasets collected by various efforts \citep{gunter2020commercially,gunter2019model,makridis2021openacc}, which result in a large ACC pool with 17 different car models from eight manufacturers.  Specifically, based on the experimental data, we have directly estimated the microscopic speed-spacing ($s-v$) relationships of the ACC systems in equilibrium conditions (in the remaining of the paper, the $s-v$ relationships all refer to the equilibrium conditions). Then, we have compared the $s-v$ relationships when the same ACC system is (a) in different positions of a platoon, (b) in different engine modes, and (c) under different time headway (hereafter headway) settings, and the relationships across different ACC car models from the same manufacturers.  After that, we have translated the $s-v$ relationships to the fundamental diagrams (flow-density relationship) and conducted a thorough analysis of the $s-v$ and FD features.

We have the following major findings: (i) a linear $s-v$ approximation fits well for the ACC systems in the medium (low) to high speed range but the $s-v$ likely differs in the very low speeds.  (ii) Headway is a key parameter for $s-v$ while vehicle position and engine mode do not change $s-v$ (pertaining to car models tested in MA experiment).  (iii) The ACC $s-v$ and FDs are very different from HDVs; i.e., at minimum headway, ACC spacing can be much smaller than HDVs and result in much larger flow, but at maximum headway, ACC spacing is often much larger than HDVs and flow is much smaller.  (iv) ACC jam spacing is much smaller than HDVs, which will reduce road storage (jam density is smaller).  (v) Some ACC systems have extremely large wave speed \RoundTwo{in the observed speed range}, which may impose safety hazard.

The remainder of  this paper is structured as follows: Section 2 introduces the data sources and Section 3 presents methodologies for the analysis. Section 4 compares the $s-v$ relationships of different ACC systems under varied settings and across different ACC system from the same manufacturer.  Section 5 further analyzes the features of $s-v$  relationships and the FDs.  Section 6 presents the conclusions and discussions.

\section{\textbf{Data}} 


The data analyzed in this paper are from three sources, the MA experiments \citep{li2021ACC}, the Work Experiments \citep{gunter2020commercially,gunter2019model}, the OpenACC experiments \citep{makridis2021openacc}, and the GA experiments. \RoundTwo{Data access will be provided through our ACC website: https://sites.google.com/view/accresources/home.}

\subsection{\textbf{MA Experiments}}
\begin{figure}[h]
\centering
\includegraphics[width=.8\textwidth]{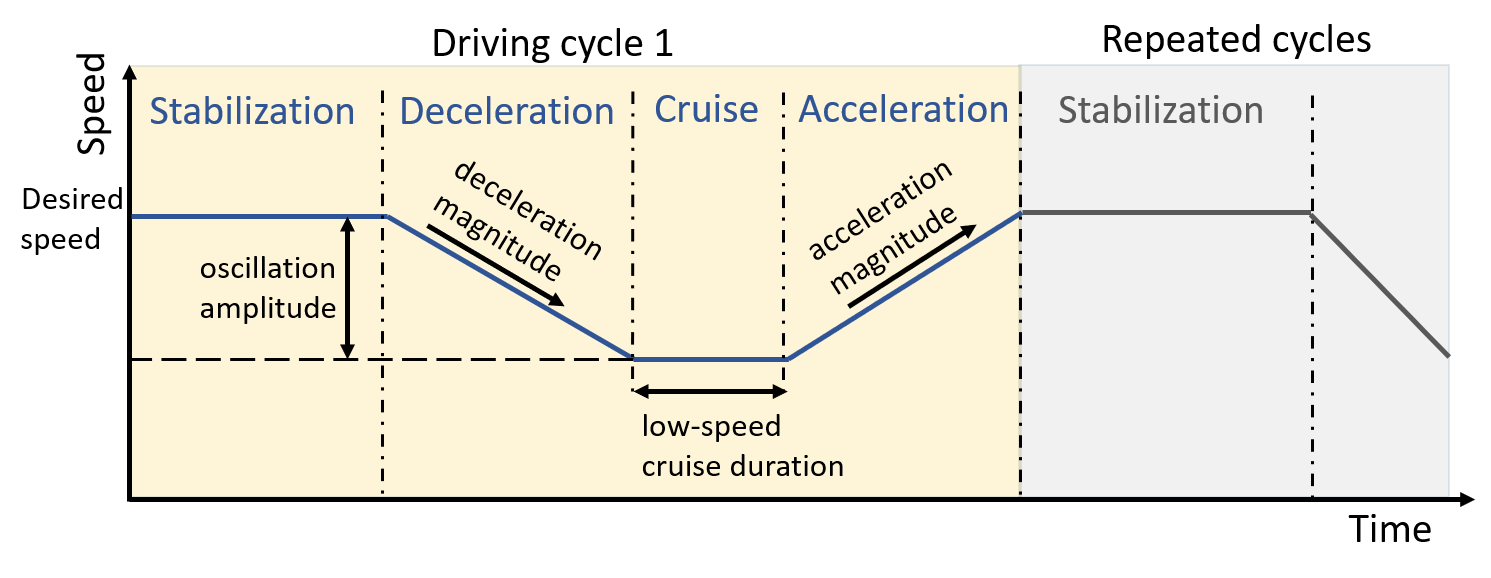}
\caption{Designed speed profile for the lead HDV in MA Experiments}
\label{speed_profile}
\end{figure}

We briefly introduce the \textit{MA experiments} (see \cite{li2021ACC} for more details).  The experiments used a three-vehicle platoon for a set of CF scenarios, where the lead vehicle was an HDV followed by two identical ACC vehicles with the same setting, referred to as ACC1 and ACC2. A high-accuracy GPS device uBlox EVK-M8T was installed on-board at the same position (upper right of the windshield) on the three cars collecting the location and velocity data. The mean location and velocity errors of the GPS device were respectively 0.89 m and 0.10 m/s based on our validation. During the experiments, the lead HDV driver was instructed to produce driving cycles that consist of steady and non-steady traffic conditions. \RoundTwo{Specifically, Figure \ref{speed_profile} shows a typical driving cycle, where the HDV first traveled at a stable speed (using cruise control), then conducted a deceleration-acceleration process, and finally  resumed the initial stable speed}. Consecutive driving cycles were separated by extensive stabilization periods (each period lasted at least 30$s$ when the speed was smaller than 20.1$m/s$ (45$mph$) and 45$s$ at higher speed).  The periods with stable speed produce the steady traffic conditions while the deceleration-acceleration (also called an oscillation) represents a disturbance.  The parameters of the driving cycles were varied to produce different combinations, including the initial stable speed and oscillation amplitude.  

Four ACC systems (i.e. car models) from three different manufacturers were tested. Their main features are listed in Table \ref{vehicle_inventory}. For each system, two headway settings, 1 and 3, were tested. Note that headway 1 is the minimum headway for all systems, while headway 3 is the maximum headway for ACC system Z, but a medium headway for systems W, X, and Y. Additionally, systems Y and Z have different engine modes \RoundTwo{(set on the dashboard as "normal" or "sport/power", indicating how aggressive the powertrain responses to the pedal command)}, which were separately tested. For systems X, Y, and Z, a total of 96 driving cycles were conducted under three speed levels (32 in each speed level) in each headway setting and engine mode. The equilibrium speed range was about \SI{15}{m/s} (\SI{35}{mph}) to \SI{29}{m/s} (\SI{65}{mph}). For system W, 64 driving cycles were conducted under two speed levels. The equilibrium speed range was about \SI{20}{m/s} (\SI{45}{mph}) to \SI{31}{m/s} (\SI{70}{mph}). 

\subsection{\textbf{Work Experiments}}

The Work Experiments \citep{gunter2020commercially,gunter2019model} used a two-vehicle setup, where an HDV leader with designed speed profiles was followed by one ACC vehicle. The trajectory data were collected using the same GPS device (uBlox EVK-M8T) as the MA Experiments. The reported mean location and velocity errors were respectively 0.43 m and 0.06 m/s. A total of eight systems were tested.  The speed ranges of equilibrium conditions vary but mostly are from \SI{15}{m/s} (\SI{35}{mph}) to \SI{33}{m/s} (\SI{75}{mph}).
The main features of the car models are listed in Table \ref{vehicle_inventory}. 
It should be noted that that system 
H is from the same manufacturer as W/X (in MA experiments).  

In the \textit{Work Experiments}, the key components of the leader speed profiles are similar to the MA experiments, including repeated stabilization period and deceleration or/and acceleration process. For each car model, two headway settings (minimum and maximum) were respectively tested with the same leader profiles.

\subsection{\textbf{OpenACC Experiments}}

A total of four experiments were conducted via the OpenACC efforts \citep{makridis2021openacc} with different setups. Specifically, the AstaZero experiment tested five ACC vehicles in a protected track which mimics a rural road environment. Minimum and maximum headway settings were tested. The leader was driven in cruise control to produce designed speed profiles similar to the MA and Work experiments. Another experiment, called ZalaZone, tested ten ACC vehicles (all from different car models) on two test tracks with similar leader design as in AstaZero. Minimum, maximum, and an additional medium headway settings were tested for each vehicle. Note that these two experiments used different measurement systems. AstaZero used a differential-GNSS level measurement solution with a precision of \SI{2}{cm/s} in speed and \SI{2}{cm} in the positioning measurement. ZalaZone fused the data from three sources (Race logic VBOX, Ublox 9 chipset, and a tracker app). The most precise source (Race logic VBOX) provided the data with \SI{2}{cm} accuracy in position and \SI{0.1}{km/h} (\SI{2.8}{cm/s}) in speed.  Besides these two experiments, there are two additional ones, Ispra-Cherasco and Ispra-VicoLungo, that tested two and five ACC vehicles, respectively, on public highway with an HDV leader producing random small perturbations. All vehicles were tested in minimum headway setting. 

For our analysis, we only use the AstaZero and ZalaZone data as the other two only have a small number of equilibrium periods, which are not enough for a meaningful analysis.  The equilibrium speed range of AstaZero is about \SI{13}{m/s} to \SI{27}{m/s} and ZalaZone is about \SI{7}{m/s} to \SI{16}{m/s}. The main features of the car models are listed in Table \ref{vehicle_inventory}. 

 \begin{table}[h]
\caption{Tested ACC car models from the three sources of experiments}
\centering
\begin{threeparttable}
\small{
\begin{tabular}{|>{\centering\arraybackslash}m{1.75cm}|>{\centering\arraybackslash}m{1.25cm}|>{\centering\arraybackslash}m{1.75cm}|>{\centering\arraybackslash}m{1.9cm}|>{\centering\arraybackslash}m{1.5cm}|>{\centering\arraybackslash}m{1.4cm}|>{\centering\arraybackslash}m{1.4cm}|>{\centering\arraybackslash}m{2.1cm}|}
\hline
manufacturer&car model & year & drive & style        & ACC speed range  & headway range & tested speed range ($m/s$) \\ \hline
\multicolumn{8}{|c|}{MA experiments$^\star$}\\\hline

3&W                   &2018   & electric       & SUV & full                   &1-7    & (20, 31)          \\\hline
3&X                   & 2018/2019$^\dagger$     & electric       & sedan & full                   &1-7    & (15, 29)           \\\hline
4&Y                   & 2019          & gasoline     & sedan        & full$^\ddag$                      &1-4     & (15, 29)   \\\hline
1&Z                   & 2018/2019$^\dagger$     & hybrid         & hatchback        & full$^\ddag$                      &1-3  & (15, 29)       \\\hline
\multicolumn{8}{|c|}{Work experiments$^\star$}\\\hline
1&A                   &2018   & gasoline       & SUV & $>$25$mph$                   &1-3    & (15, 33)          \\\hline
1&B                   & 2018    & gasoline       & sedan & $>$25$mph$                   &1-3    & (15, 33)          \\\hline
1&C                   & 2018          & hybrid     & hatchback        & $>$25$mph$                      &1-3     & (15, 33) \\\hline
1&D                   & 2018     & gasoline         & SUV        & $>$25$mph$                      &1-3  & (15, 33)     \\\hline
2&E                   & 2018    & gasoline         & SUV        & $>$25$mph$                      &1-3  & (15, 33)     \\\hline
2&F                   & 2018     & gasoline         & SUV        & $>$25$mph$                      &1-3  & (15, 33)     \\\hline
2&G                   & 2018     & gasoline         & SUV        & $>$25$mph$                      &1-3  & (15, 33)     \\\hline
3&H                   & 2015     & electric         & sedan        & full                      &1-7  &(5, 31)     \\\hline
\multicolumn{8}{|c|}{OpenACC experiments - AstaZero}\\\hline
Tesla&Model 3                   &2019   & electric       & sedan & full                  &1-7    & (13, 27)          \\\hline
Audi&A6                   &2018   & diesel       & sedan &         full          &   1-4 & (13, 27)          \\\hline
BMW&X5                   &2018   & diesel       & SUV &            full       &   1-4 & (13, 27)          \\\hline
Mercedes&A Class                   &2019   & gasoline       & sedan &         $>$15$mph$          &   1-4 & (13, 27)          \\\hline

\multicolumn{8}{|c|}{OpenACC experiments - ZalazONE}\\\hline
Tesla&Model 3                   &2019   & electric       & sedan & full                  &1-7    & (7, 16)          \\\hline
Tesla&Model X                   &2016   & electric       & SUV &         full          &   1-7 & (7, 16)          \\\hline
Tesla&Model S                   &2018   & electric       & sedan &         full          &   1-7 & (7, 16)          \\\hline
Audi&A4                  &2019   & gasoline       & sedan &          full         &   1-4 & (7, 16)          \\\hline
Audi&E-tron                   &2019   & electric       & SUV &          full         &   1-4 & (7, 16)          \\\hline
BMW&I3 s                   &2018   & electric       & hatchback &          full         &   1-4 & (7, 16)          \\\hline
Mercedes&GLE450                   &2018   & gasoline       & SUV &          $>$15$mph$          &   1-4 & (7, 16)          \\\hline
Jaguar& I-Pace                   &2019   & electric       & SUV &          full         &   1-3 & (7, 16)          \\\hline
Toyota&RAV4                   &2019   & gasoline       & SUV &          $>$25$mph$         &   1-3 & (7, 16)          \\\hline
Mazda&Mazda3                 &2019   & gasoline       & sedan &         $>$20$mph$           &  1-4  & (7, 16)          \\\hline
\multicolumn{8}{|c|}{GA experiments$^\star$}\\\hline
3&X                   & 2018     & electric       & sedan & full                   &1-7    & (0, 33)          \\\hline

\end{tabular}
}
\vspace{-12pt}
\begin{tablenotes}[flushleft]
    \footnotesize  
    \item $\star$The car models are anonymous in the MA, Work, and GA experiment. 
    \item $\dagger$Constrained by the availability, the two tested cars for model X and Z are manufactured in two continuous years. But they are ensured to be the same generation and with the same equipment.
    \item $\ddag$The ACC function can only be activated above 25$mph$. Once activated, it keeps on at lower speed until fully stop.
\end{tablenotes}
\end{threeparttable}
\label{vehicle_inventory}
\end{table}

\subsection{\textbf{GA Experiments}}

The GA experiments were conducted by the research team particularly focusing on the ACC car model X.  The measurement device was the same as the one used in the MA experiments. The tests were conducted on an urban arterial with a two-vehicle platoon setup. The leader was driven in cruise control in multiple speed levels and the following vehicle always engaged with ACC.  Moreover, when the leader stopped at intersections, the follower ACC was still engaged in the ACC mode, which provides a direct measurement of the jam spacing for the model X.




\section{\textbf{Methodologies}}
This section introduces the method to estimate the $s-v$ relationship based on the experiment data and the statistical approach to compare two different $s-v$ relationships.

\subsection{\textbf{Estimating the $s-v$ relationship}}
To estimate the $s-v$ relationship, we first identify the equilibrium intervals to measure the speed and spacing, which are then used to estimate a $s-v$ relationship using a first-order approximation.  In the estimation, we also have to make sure that the measurements fall into different speed levels (captured by speed bins\RoundTwo{, see more details in Section \ref{s-v relationship}}) to produce a meaningful projection.  

\subsubsection{\textbf{Identification of Traffic States in Equilibrium}}\label{equilibrium_search_section}

To estimate the $s-v$ relationship, we first need to identify traffic states in equilibrium.  From the vehicle trajectories, we recognize the equilibrium intervals based on five strict thresholds: (i) steady speed, i.e., both the leader and follower speed variation (maximum - minimum) does not exceed \SI{0.45}{m/s} (\SI{1}{mph}) for at least \SI{10}{s}, (ii) speed difference between the leader and follower is not larger than \SI{0.45}{m/s} in the same interval; (iii) stable spacing, i.e., spacing variation does not exceed \SI{1}{m} in the same interval; (iv) moderate grade, i.e., the roadway grade is not larger than $3\%$ (absolute value) in the same interval and for \SI{10}{s} before it; and (v) far apart from disturbances, i.e., the interval is at least \SI{10}{s} after the end of the previous driving cycle. \RoundTwo{In general, thresholds (i)-(iv) are sufficient, though threshold (v) adds another layer to make sure that the extracted segment is not impacted by the disturbance residual. We have done the comparison with and without (v) using MA experiments and found marginal difference in the outcome.  In the following analysis, we apply (i)-(v) for MA experiments but (i)-(iv) for other experiments to reserve a larger sample. } Here we used the wavelet transform algorithm per \cite{zheng2011applications} to identify start and end of the driving cycles. Figure \ref{equilibrium_example} shows an example of the recognized interval in equilibrium. \RoundTwo{Note that if the data in general has large fluctuations in the traffic states, it may be very difficult to find enough equilibrium intervals.  In this case, one may have to relax the thresholds and consider additional criteria such as acceleration, depending on the features of the data.}

\noindent
\begin{minipage}{\linewidth}
\centering
\includegraphics[width=.65\textwidth]{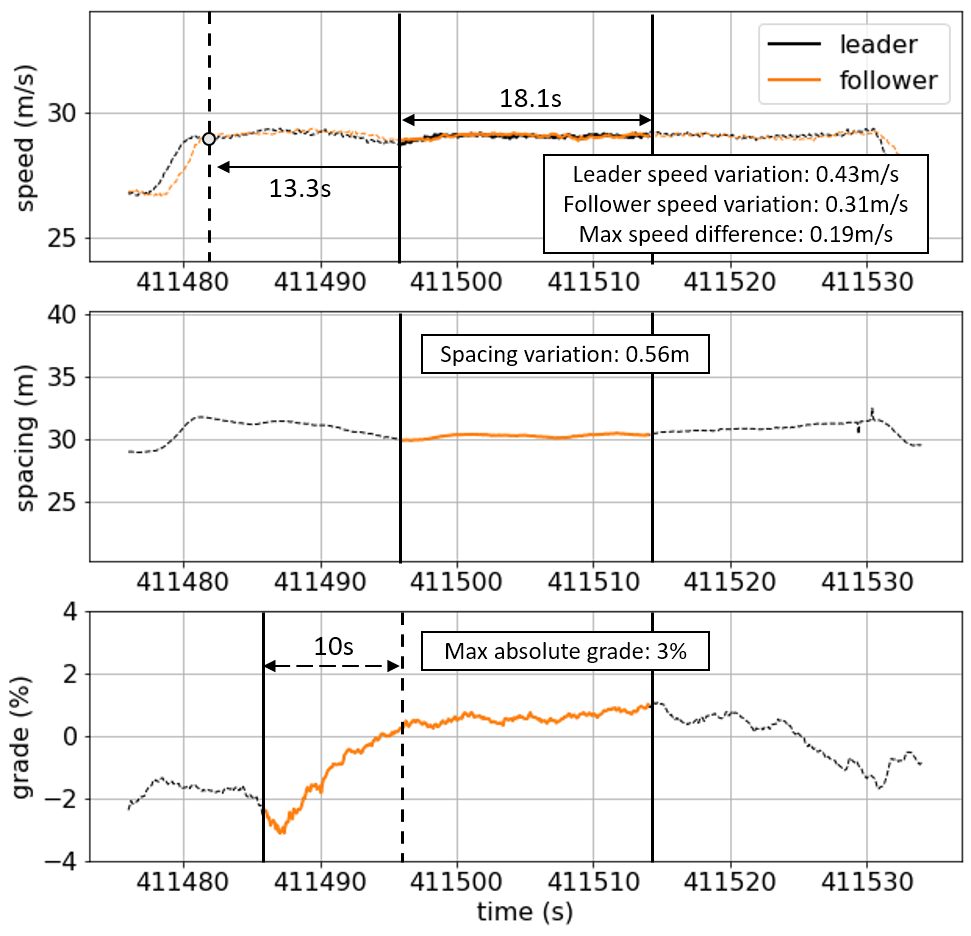}
\captionof{figure}{An example of an equilibrium interval (ACC system X)}
\label{equilibrium_example}
\end{minipage}

\subsubsection{\textbf{Estimation of the speed-spacing relationship of ACC systems}}\label{s-v relationship}

Based on the equilibrium intervals, we can estimate the $s-v$ relationship of an ACC system.  Specifically, for each interval $i$, we measure the average speed and spacing, denoted as ($v_i$, $s_i$).   Then we estimate the $s-v$ relationship using a first-order formulation. Here linearity is assumed as the literature has shown that it is a good approximation for the commercial ACC systems tested \citep{gunter2020commercially} and also widely used in design of ACC controllers \citep{naus2010string,milanes2013cooperative,zhou2021impact}. The spacing-speed relationship is described as:
\begin{equation}\label{linear SV}
    s = \tau_0 v + \delta_0 ,
\end{equation} 
where $s$ is the equilibrium spacing at speed $v$, $\tau_0$ and $\delta_0$ are the headway (also referred to as the time-gap) and the constant coefficient \RoundTwo{(jam spacing if the $s-v$ relationship holds at stop)}, respectively.  Here $s$ and $v$ are directly measured from the data and $\tau_0$ and $\delta_0$ are to be estimated. \RoundTwo{We also note that here the spacing $s$ is the front to front bumper distance, which is directly provided in the MA, GA, and Work dataset. On the other hand, the OpenACC dataset only provides the IVS (inter-vehicle spacing) measurement. Therefore, the spacing for OpenACC data is calculated by IVS adding the leader vehicle length (also provided in the dataset).}

To produce a meaningful  $s-v$  approximation, we require that the measurements should have a good representation of different speed levels.  Specifically, the measurements are classified into different speed bins.   \RoundTwo{Each equilibrium interval results in one data point (speed and spacing), which takes the average (follower) speed and average spacing across that equilibrium interval. In the bin clustering, we require that (i) the speed range of a bin (maximum speed minus minimum speed of all data points in one bin) does not exceed 2m/s, and (ii) the speed difference between one data point and its nearest neighbor point (i.e., the speed level is closest) in the bin should not exceed 0.5m/s.  (i) and (ii) together assure that the data points in the same bin maintain a similar speed level. For one ACC setting, we require at least three bins to construct the  $s-v$ relationship\footnote{\RoundTwo{Note that two bins can produce a linear line but it's hard to tell how well the linear line fits the actual $s-v$ relationship. }}.}


\subsection{\textbf{Comparison of Two Different Spacing-speed Relationships}}

With the estimated $s-v$ relationships for the ACC systems, we can see whether the relationship is different between two systems (or two settings of the same system).  This is boiled down to comparing whether the estimated $\tau_0$ and $\delta_0$ are different from one system to another. To answer this question, we use the method introduced in \cite{wooldridge2016introductory}. Below we briefly introduce this method.

For two ACC systems $K$ and $J$ to be compared, their respective $s-v$ formulations are below:
\begin{equation}\label{k_eq}
    s = \tau_0^K v + \delta_0^K ~,
\end{equation}
\begin{equation}\label{j_eq}
    s = \tau_0^J v + \delta_0^J ~.
\end{equation}

\noindent The questions to be answered are: (i) whether $\tau_0^K$ is significantly different from $\tau_0^J$, and (ii) whether $\delta_0^K$ is significantly different from $\delta_0^K$. To conduct the comparison, we first combine the sample sets of the two systems, $S^K$ and $S^J$. Additionally, a dummy variable $c_J$ is added, which is set to 1 for the sample belonging to $S^J$ and 0 for the sample belonging to $S^K$. Then, we formulate a new equation as:
\begin{equation}\label{combined_eq}
    \RoundTwo{s = \tau_0^K v + \delta_0^K + \tau_0^{J-K} v \cdot c_J + \delta_0^{J-K} c_J ~,}
\end{equation}
where $v \cdot c_J$ is the interaction term of $v$ and $c_J$, which is set to the original $v_i$ (i.e., $c_J=1$) values for the samples in $S^J$ and 0 (i.e., $c_J=0$) for the samples in $S^K$ .

\RoundTwo{The two terms, $v \cdot c_J$ and $c_J$, account for the differences between the system K and J. The corresponding coefficient $\tau_0^{J-K}$ indicates the difference in slope between K and J (i.e., whether $\tau_0^J$ is significantly different from $\tau_0^K$), and $\delta_0^{J-K}$ is the intercept difference between K and J (i.e., whether $\delta_0^J$ is significantly different from $\delta_0^K$).}
Therefore, we address questions (i) and (ii) by conducting a linear regression on Equation (\ref{combined_eq}) using the combined sample set to estimate the coefficients for the variables $(v, v \cdot c_J, c_J)$. If the coefficient of the interaction term $v \cdot c_J$  is significant different from zero, $\tau_0^J$ is significantly different from $\tau_0^K$. Similarly, if the coefficient of $c_J$ is significant, $\delta_0^J$ is significantly different from $\delta_0^K$.

\section{\textbf{Comparison of ACC systems}}
This section presents a comparison of the $s-v$ relationships.  Our analysis shows that a linear relationship well fits the ACC systems in the tested speed ranges.  In fact, 75\% of the regressions have $R^2$ above 0.8, and 54\% of them are above 0.9.  Based on that, we will first compare the $s-v$ relationships in the same ACC systems when the settings vary and then compare the relationships across different ACC systems from the same manufacturer. 

\subsection{\textbf{The same ACC system with varying settings}}

Here we compare the $s-v$ relationships in the same ACC systems when the ACC vehicles are  (i) in different positions of a platoon,  (ii) in different engine modes, and (iii) under different headway settings. The analysis is based on the data from the \textit{MA Experiments} involving four car models.  In the experiment the impact of a parameter setting was tested with other factors controlled.  Therefore, the equilibrium measurements are from the same traffic conditions and comparable.  Other experiments do not have such controlled tests. 


To see the impact of vehicle position in a platoon, We estimate one $s-v$ relationship corresponding to each position and then compare the $s-v$ relationships across different positions.  Similar approach is used to see the difference with different engine mode and headway settings.  We have observed the following remarks: 

\textbf{Remark 1:} the $s-v$ relationship remains the same when the same ACC car model is in different position of a platoon. Specifically, recall that the MA Experiments had a three-vehicle platoon setup where two identical ACC vehicles (always with the same setting) followed an HDV.  That allows us to estimate the $s-v$ for ACC1 and ACC2; see results in Table \ref{compare_same_system}  in the appendix. One can see that the differences are not significant (95\% significance level is assumed for the entire study) regardless of the headway setting and engine mode, and that result holds for all four car models we tested.  Based on the results here, we combine the data from ACC1 and ACC2 in our following comparisons regarding other aspects.

\textbf{Remark 2:} the $s-v$ relationship remains unchanged in different engine modes; see Table \ref{compare_engine} in the appendix. In the MA experiments, two ACC systems (Y and Z) had two different modes of operation, normal and sport/power, which were tested with other parameters controlled.  The change of parameters in different mode is not significant for either ACC system.  Based on the results here, we combine the data from different engine modes in our following comparisons regarding other aspects.

\textbf{Remark 3:} when an ACC system changes the headway setting, $\tau_0$ changes too (larger values in larger headway) but $\delta_0$ does not necessarily change; see Table \ref{compare_headway}. Notably, the comparison here expand from the MA Experiment to the four experiments, with 17 systems in total (some are not included due to insufficient speed coverage). 
Note that several car models in ZalaZone were tested in three levels of headway setting, in which case the minimum and maximum are used for the comparison. 

Interestingly, in 10 out of the 16 systems, $\delta_0$ remains similar as headway increases. In four systems, $\delta_0$ decreases as headway increases. For example, for ACC system F, $\tau_0 =0.77s$ and $\delta_0=13.24m$ for the minimum headway setting, but in the maximum headway $\tau_0$ increases to \SI{2.02}{s} and $\delta_0$ decreases to \SI{4.51}{m}. In two systems (E and X), $\delta_0$ increases as headway increases, which seems surprising.  
These results suggest that ACC may adjust both parameters as headway setting changes. Overall, spacing is smaller as headway decreases.  Some ACC systems use a larger $\delta_0$ to compensate the spacing drop due to headway decrease but that does not reverse the decreasing trend of the net spacing. 

The three remarks obtained suggest that, headway is a key parameter that affects the equilibrium.  When we consider different ACC systems, the comparison should be done for each headway setting.  

\begin{table}[!h]
\caption{ACC under different headway settings}\label{compare_headway}
\centering
\begin{threeparttable}
\footnotesize{
\begin{tabular}{|>{\centering\arraybackslash}m{2cm}|>{\centering\arraybackslash}m{1cm}|>{\centering\arraybackslash}m{1cm}|>{\centering\arraybackslash}m{1.1cm}|>{\centering\arraybackslash}m{1cm}|>{\centering\arraybackslash}m{1cm}|>{\centering\arraybackslash}m{1.1cm}|c|c|}
\hline
\multirow{2}{*}{System} & \multicolumn{3}{c|}{Small headway}   & \multicolumn{3}{c|}{Large headway}   & \multirow{2}{1.9cm}{\centering $\tau_0$ difference (p-value)} & \multirow{2}{1.9cm}{\centering $\delta_0$ difference (p-value)} \\ \cline{2-8}
                        & $\tau_0$ (s) & $\delta_0$ (m) & sample  & $\tau_0$ (s) & $\delta_0$ (m) & sample  &                                     &                                     \\ \hline
A & 0.89 & 12.39 & 68  & 1.96 & 14.83 & 43  & 1.07 (\smallp\sig) & 2.43 (0.08)  \\ \hline
B & 0.87 & 10.92  & 59  & 2.10 & 10.83 & 47  & 1.23 (\smallp\sig) & -0.10 (0.98)   \\ \hline
C & 0.87 & 14.43 & 45  & 2.25 & 10.36 & 50  & 1.41 (\smallp\sig) & -4.07 (0.25)  \\ \hline
D & 0.78 & 13.48 & 46  & 1.98 & 14.09 & 52  & 1.20 (\smallp\sig) & 0.61 (0.85)   \\ \hline
E & 1.27 & 4.97  & 54  & 2.02 & 9.00  & 52  & 0.76 (\smallp\sig) & 4.02 (0.02\sig)   \\ \hline
F & 0.77 & 13.24 & 57  & 2.02 & 4.51  & 55  & 1.24 (\smallp\sig) & -8.74 (\smallp\sig)  \\ \hline
G & 0.61 & 17.44 & 54  & 2.00 & 5.36  & 51  & 1.39 (\smallp\sig) & -12.07 (\smallp\sig) \\ \hline
H & 0.60 & 11.78 & 63  & 1.84 & 9.85  & 46  & 1.24 (\smallp\sig) & -1.93 (0.01\sig)  \\ \hline
W & 0.57 & 14.18 & 92  & 0.93 & 15.33 & 56  & 0.36 (\smallp\sig) & 1.15 (0.57)  \\ \hline
X & 0.47 & 16.86 & 78 & 0.84 & 16.78 & 27  & 0.37 (\smallp\sig) & -0.08 (0.96)  \\ \hline
Y & 1.10 & 4.56  & 30  & 2.07 & 5.14  & 45  & 0.97 (\smallp\sig) & 0.58(0.70)  \\ \hline
Z & 0.87 & 11.93 & 28  & 1.98 & 18.25 & 28  & 1.11 (\smallp\sig) & 6.33 (0.10)  \\ \hline

Model 3       & 0.47 & 15.54 & 96  & 2.21 & 11.27 & 119  & 1.74 (\smallp\sig) & -4.26 (\smallp\sig)  \\ \hline
Model X       & 0.50 & 11.42 & 45  & 1.18 & 15.65 & 99  & 0.68 (\smallp\sig) & 4.22 (0.01\sig)  \\ \hline
Model S       & 1.03 & 9.72 & 81  & 2.39 & 9.63 & 89  & 1.36 (\smallp\sig) & -0.09 (0.91)  \\ \hline
I3 s & 1.33 & 7.86 & 100  & 2.15 & 7.78 & 113  & 0.82 (\smallp\sig) & -0.08 (0.97)  \\ \hline

\end{tabular}
}
\vspace{-10pt}
\begin{tablenotes}[flushleft]
    \footnotesize  
    \item \sig indicate significant difference at 95\% confidence level. 
\end{tablenotes}
\end{threeparttable}
\end{table}

\subsection{\textbf{Different ACC systems from the same manufacturer}}
Here we compare the $s-v$ relationship across different car models from the same manufacturer. We have analyzed the car models from the anonymous Manufacturers 1 and 2 in the Work experiments, Manufacturer 3 in the MA experiments, and Tesla in the OpenACC data.

For each car model, we have estimated one $s-v$ relationship for each headway setting and our comparison is conducted under the same headway setting.  We caution that, the true $s-v$ relationship may not be a single linear relationship for the entire speed range.  For example, it could be piece-wise linear at different speed ranges or even non-linear. To have fair comparisons, our comparison is applied to car models tested in the same experiment that has similar traffic conditions (e.g., same speed range and same geometric conditions) and the same measurement system (thus same error resolution). 
Note that some car systems were tested in more than one experiments and we find that the comparison results are generally consistent. 

The results show that in the four manufacturers examined, two (Manufacturers 1 and 3) have the same $s-v$ relationship for the different car models; see Figure \ref{same_manufacturer_comparison} (a) for the four car models (A-D) of Manufacturers 1 from the Work Experiment, and Figure \ref{same_manufacturer_comparison} (c) for the two car models of Manufacturer 3  (W and X) in the MA experiments. 


For the other two manufacturers (Manufacturer 2 and Tesla), each of them has three different car models tested, and the ACC systems show significant differences; see Table \ref{compare_Manufacturer}. 
Specifically, the three car models from Manufacturers 2 are different at minimum headway setting but seem similar in maximum headway.  The three Tesla models tested in the OpenACC experiment seem to differ from each other in general, except for Model S and Model 3 at maximum headway. Notably, one can see the spacing variation under the same speed could be significant in many cases, such as Figure \ref{same_manufacturer_comparison}(d) in which the variance even overshadows the difference of mean between the systems.



The result that the $s-v$ relationship varies with car models seems unexpected.  We conjecture that the differences are related to vehicle mechanical features (e.g., power-to-weight ratio and power-train types), hardware generation, and the design and calibration of the low-level controller. 
This is being investigated by the research team.  

Notably, since the car models in all the experiments were not tested in the full speed ranges, we caution that future research is needed to generalize the results in this section to the full speed range.

\begin{table}[H]
\caption{Different car models from the same manufacturer}\label{compare_Manufacturer}
\centering
\begin{threeparttable}
\footnotesize{
\begin{tabular}{|>{\centering\arraybackslash}m{1.25cm}|>{\centering\arraybackslash}m{1.1cm}|>{\centering\arraybackslash}m{1.1cm}|>{\centering\arraybackslash}m{1.1cm}|>{\centering\arraybackslash}m{1.1cm}||>{\centering\arraybackslash}m{1.3cm}|>{\centering\arraybackslash}m{1.1cm}|>{\centering\arraybackslash}m{1.1cm}|>{\centering\arraybackslash}m{1.1cm}|>{\centering\arraybackslash}m{1.1cm}|>{\centering\arraybackslash}m{1.1cm}|}

\hline
\multicolumn{10}{|c|}{Manufacturers 2 (Work experiments)}\\
\hline
    \multirow{3}{1.25cm}{Minimum headway} & \multicolumn{2}{c|}{F}    & \multicolumn{2}{c||}{G}     & \multirow{3}{1.3cm}{Maximum headway} & \multicolumn{2}{c|}{F} & \multicolumn{2}{c|}{G}     \\ \cline{2-5} \cline{7-10} 
    
     & \multicolumn{2}{c|}{difference (p-value)}    & \multicolumn{2}{c||}{difference (p-value)}     &  & \multicolumn{2}{c|}{difference (p-value)} & \multicolumn{2}{c|}{difference (p-value)}     \\ \cline{2-5} \cline{7-10} 
                                 & $\tau_0$        & $\delta_0$          & $\tau_0$         & $\delta_0$          &                            & $\tau_0$        & $\delta_0$       & $\tau_0$         & $\delta_0$          \\ \hline
                                 
E                                & -0.44 (\smallp\sig) & 6.75 (\smallp\sig) & -0.59 (\smallp\sig) & 10.56 (\smallp\sig) & E                                & -0.10 (0.14) & -3.01 (0.06) & 0.00 (0.98) & -4.55 (\smallp\sig) \\ \hline

F                                & \multicolumn{2}{c|}{-}     & -0.15 (\smallp\sig) & 3.81 (\smallp\sig)  & F                                & \multicolumn{2}{c|}{-}      & 0.10 (0.07) & -1.54 (0.22) \\ \hline

                                 

\multicolumn{10}{|c|}{Tesla (OpenACC experiments)}\\
\hline
    \multirow{3}{1.25cm}{Minimum headway} & \multicolumn{2}{c|}{Model X}    & \multicolumn{2}{c||}{Model S}     & \multirow{3}{1.3cm}{Maximum headway} & \multicolumn{2}{c|}{Model X} & \multicolumn{2}{c|}{Model S}     \\ \cline{2-5} \cline{7-10} 
    
     & \multicolumn{2}{c|}{difference (p-value)}    & \multicolumn{2}{c||}{difference (p-value)}     &  & \multicolumn{2}{c|}{difference (p-value)} & \multicolumn{2}{c|}{difference (p-value)}     \\ \cline{2-5} \cline{7-10} 
                                 & $\tau_0$        & $\delta_0$          & $\tau_0$         & $\delta_0$          &                            & $\tau_0$        & $\delta_0$       & $\tau_0$         & $\delta_0$          \\ \hline
                                 
Model 3                               & 0.04 (0.83) & -4.11 (0.01\sig) & 0.57 (\smallp\sig) & -5.82 (\smallp\sig) & Model 3                               & -1.03 (\smallp\sig) & 4.38 (\smallp\sig) & 0.18 (0.10) & -1.65 (0.10) \\ \hline

Model X                                & \multicolumn{2}{c|}{-}     & 0.53 (\smallp\sig) & -1.71 (0.21)  & Model X                                & \multicolumn{2}{c|}{-}      & 1.21 (\smallp\sig) & -6.02 (\smallp\sig) \\ \hline

\end{tabular}
}

\end{threeparttable}
\end{table}

\noindent
\begin{minipage}{\linewidth}
\centering
\begin{minipage}[b]{.325\linewidth}
\centering
\footnotesize
\includegraphics[width=\textwidth]{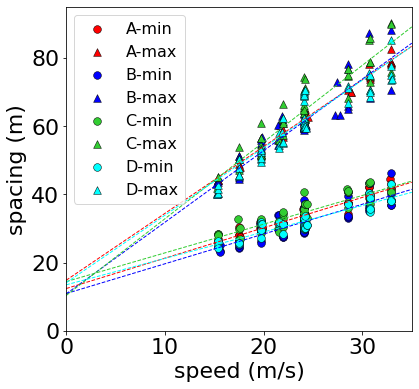}
\\[-2pt]
(a) Manufacture 1
\end{minipage} 
\hspace{30pt}
\begin{minipage}[b]{.325\linewidth}
\centering
\footnotesize
\includegraphics[width=\textwidth]{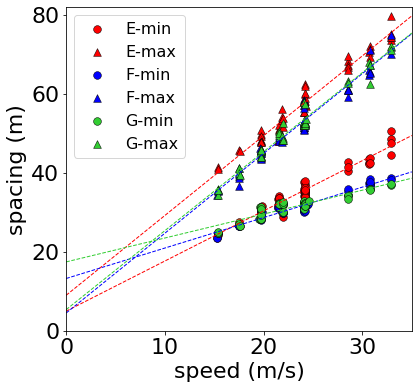}
\\[-2pt]
(b) Manufacture 2
\end{minipage}

\begin{minipage}[b]{.325\linewidth}
\centering
\footnotesize
\includegraphics[width=\textwidth]{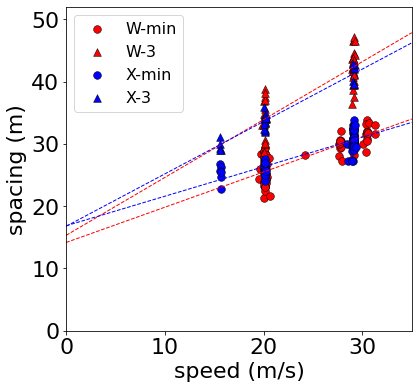}
\\[-2pt]
(c) Manufacture 3
\end{minipage} 
\hspace{30pt}
\begin{minipage}[b]{.325\linewidth}
\centering
\footnotesize
\includegraphics[width=\textwidth]{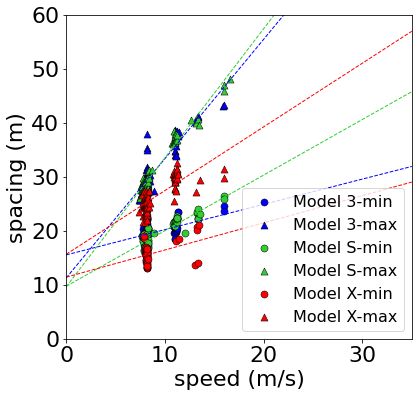}
\\[-2pt]
(d) Tesla
\end{minipage}

\captionof{figure}{$s-v$ relationships of ACC systems from the same manufactures}\label{same_manufacturer_comparison}
\end{minipage}

\section{\textbf{Characteristics of ACC $s-v$ relationships and the fundamental diagrams}}\label{FD}
In this section, we investigate the characteristics of the $s-v$ relationships and the corresponding fundamental diagrams (FDs) that connect the flow, density, and speed from the macroscopic perspective. 

We consolidate the data across the four experiments to obtain a more comprehensive coverage of the  $s-v$ relationships whenever possible.  Specifically, some car models were tested in more than one experiments. In that case, we conduct a statistical test on the estimated $s-v$ relationships from different data sources and combine them only if the estimated results are statistically insignificant. We find that the estimated $s-v$ relationships from different experiments are consistent, except for one case - one car model estimated from AstaZero differed from the estimations from MA experiment and ZalaZone.  That is likely because the measurement systems have different measurement errors (more discussions on this will follow in Session \ref{conclusion}).  Nevertheless, for this case, we have excluded the data from AstaZero and combine the data from the other two sources.   

Using the consolidated data, we first estimate $s-v$ relationships for each car model (one $s-v$ for each headway setting).  After that, we estimate the corresponding flow-density ($q-k$) relationship; i.e., the fundamental diagram.  Specifically, a linear $s-v$ relationship translates to a triangular shaped FD with wave speed $w= -\delta_0/\tau_0$, jam density $k_j= 1/\delta_0$, and capacity $Cap = u_f / (u_f \tau_0 + \delta_0) =- k_j w u_f/(u_f-w)$. Note that the capacity depends on free-flow speed $u_f$. Here we project the capacity at $u_f=105km/h$ $(65mph)$ as it is near the higher bound of the ACC systems tested. \RoundTwo{Notably, the capacity obtained is the \textit{equilibrium capacity}, which may differ from the actual capacity one may observed from loop detectors on the road.  More discussion is provided in Section 6.}


\subsection{\textbf{A typical $s-v$ and FD}}
Here we first show a typical example of the estimated $s-v$ and FD.

Figure \ref{sv_qk_single} shows the results of ACC system H. One can see that the ACC equilibrium points are well aligned on the linear curves in both headway settings. The linear relationship holds in a wide speed range, from low to high speed (\SI{7}{m/s}-\SI{30}{m/s}). In the plot (and other $s-v$ plots later), we provide the references from HDVs (red dashed line for the mean, and orange dotted lines for the 15th and 85th percentiles, denoted as $P_{15}$ and $P_{85}$, representing HDV lower and upper bounds).  The HDV references are taken from \cite{chen2012behavioral} that measured $\tau_0$ and $\delta_0$ of individual drivers in the NGSIM dataset \citep{NGSIM_2006}. Note that, the variation of $\tau_0$ and $\delta_0$ in HDVs \citep{chen2012behavioral} stems from the difference of individual HDVs, which is a holistic outcome of the heterogeneity from drivers and the vehicles used.  Nevertheless, their distributions provide some references of the HDVs.

The ACC system is very different from HDVs, especially in minimum headway. Clearly, the H-min $s-v$ line (This paper uses H-min to represent ACC system H under \RoundTwo{minimum} headway. The same structure applies to other car models.), deviates significantly from the HDV mean. It falls outside the HDV bounds in two regions: (i) in mid-high speed levels (larger than \SI{12}{m/s}), the ACC  spacing is much smaller than HDV lower bound, translating to a much larger flow at a given speed (see Figure \ref{sv_qk_single}(b)).  Particularly, the capacity is 75\% higher than the HDV average.  (ii) In the very low speed range (close to stop), the ACC spacing is above the HDV upper bound, suggesting a much larger jam spacing ($\delta_0 =11.78m$) and smaller jam density ($k_j=85veh/km$). Moreover, one can see from Figure \ref{sv_qk_single}(b) that H-min embeds a very fast propagating wave speed (\SI{-70.7}{km/h}) that is very different from HDVs.  The H-max $s-v$ line is mostly above the HDV mean reference but is mostly inside the HDV bounds. The capacity is 20\% lower than that of the HDV mean. The projected jam spacing is still larger than HDVs ($\delta_0 =9.85m$), but the wave speed seems comparable to HDVs ($w=-5.35m/s$). 

\noindent
\begin{minipage}{\linewidth}
\centering
\begin{minipage}[b]{.49\linewidth}
\centering
\small
\includegraphics[height=175pt]{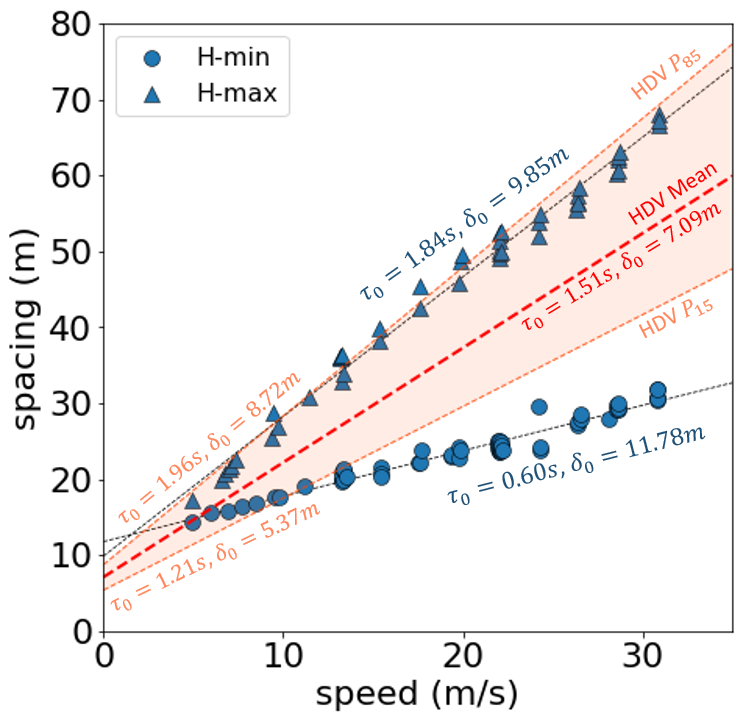}
\\[-6pt]
(a) $s-v$ relationship
\end{minipage}
\begin{minipage}[b]{.49\linewidth}
\centering
\small
\includegraphics[height=175pt]{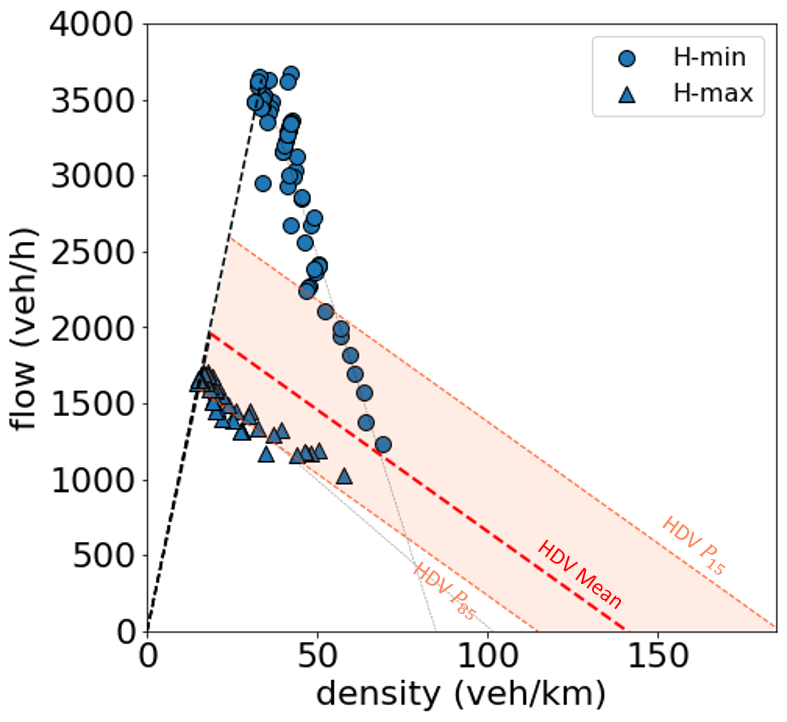}
\\[-6pt]
(b) $q-k$ relationship
\end{minipage} 
\captionof{figure}{FD of ACC system H}
\label{sv_qk_single}
\end{minipage}

\subsection{\textbf{The ACC pool tested}}
Our observations above pertaining to ACC system H are very common in the ACC pool.  Figure \ref{q_k_overall} shows the $s-v$ plots and corresponding FDs for the entire ACC pool.  The color scatters highlight systems with consolidated data and wider speed coverage.  The plots for all the ACC car models are provided in Figure \ref{sv_Manufacturer} and Figure \ref{sv_Manufacturer_other} and the estimated $s-v$ relationships are summarized in Table \ref{full_sv_regression_table} in the Appendix. In general,  a linear estimation yields a very good fit ($R^2$ is mostly above 0.8) for most of the models in the tested speed ranges, which usually covers medium, or even low, to high speeds.  Clearly, the $s-v$ is usually below the HDV average under minimum headway, translating to much larger flow and thus the capacity, but it is usually above the HDV average for maximum headway.

\noindent
\begin{minipage}{\linewidth}
\centering
\begin{minipage}[b]{.49\linewidth}
\centering
\small
\includegraphics[height=150pt]{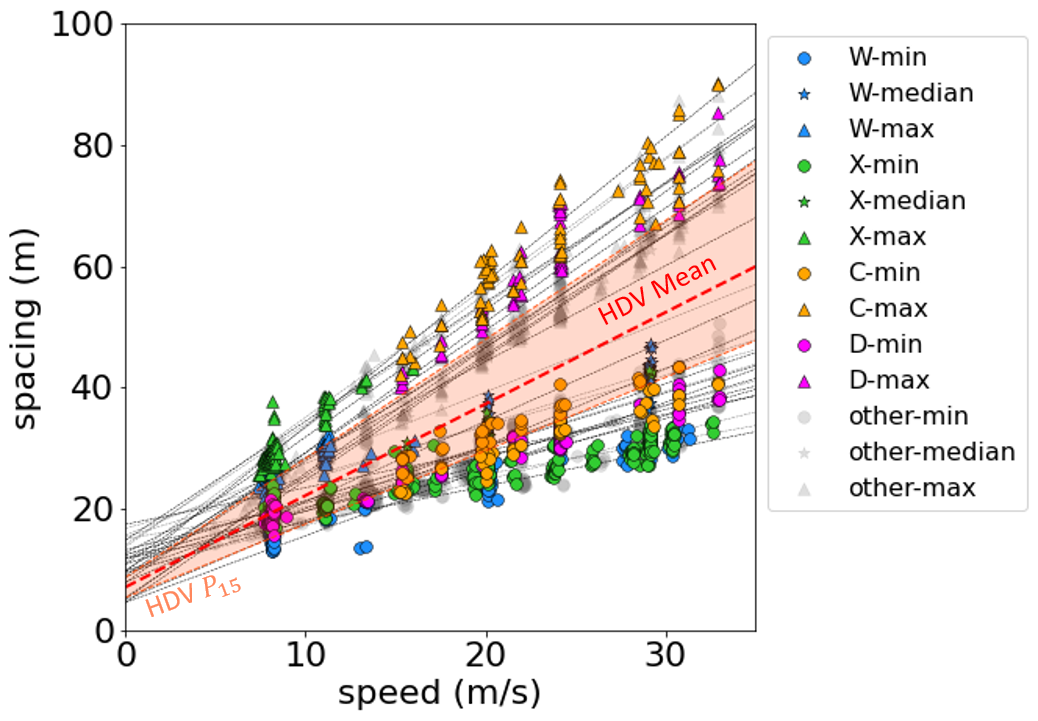}
\\[-6pt]
(a) s-v relationship
\end{minipage}
\begin{minipage}[b]{.49\linewidth}
\centering
\small
\includegraphics[height=150pt]{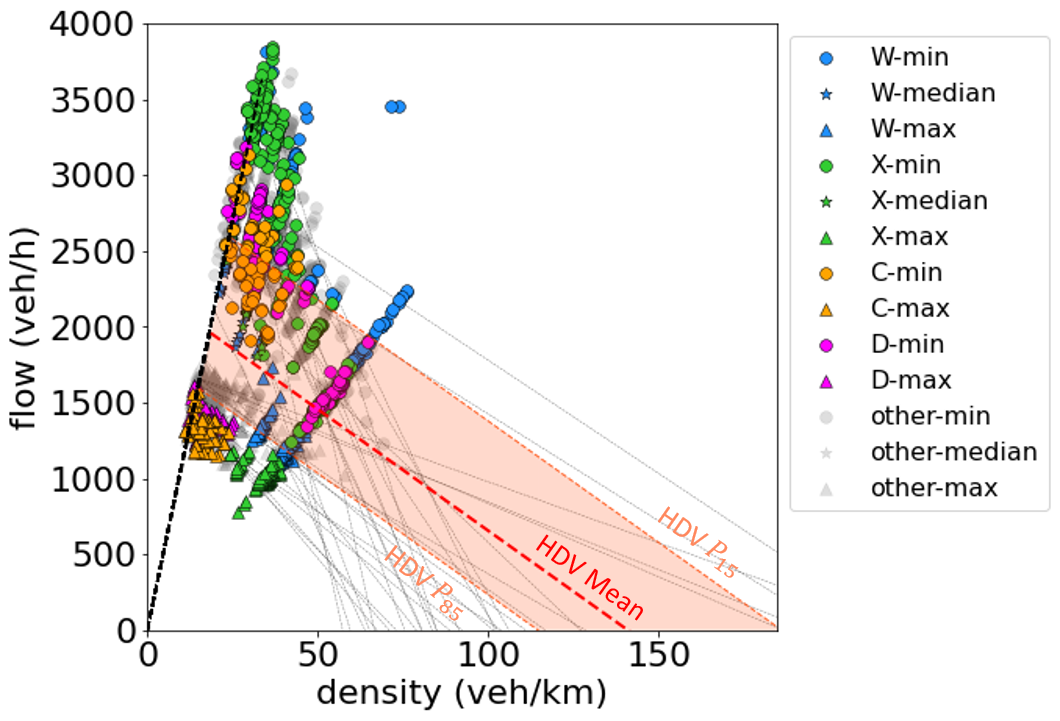}
\\[-6pt]
(b) q-k relationship
\end{minipage} 
\captionof{figure}{FD of the evaluated ACC systems (colored scatters are from the consolidated data and the grey ones from the non-consolidated ones)
}
\label{q_k_overall}
\end{minipage}

Figure \ref{time-gap-jam-spacing_dist} shows the estimated $\tau_0$ and $\delta_0$ distribution of the ACC pool examined compared to the reference - HDVs. One can see that, the distribution of $\tau_0$ ($\delta_0$) for ACC exhibits two clusters, representing the lower and upper bounds of the systems evaluated, corresponding to the minimum and maximum headway. Notice that the $\tau_0$ of the ACC is distributed in a much larger range and has a much smaller lower bound (\SI{0.54}{s} vs. \SI{0.91}{s}).  In fact, 87\% of the car models in minimum headway have $\tau_0$ smaller than HDV $P_{15}$.  Regarding $\delta_0$, the ACC pool is distributed in a much wider range and has a much larger upper bound than HDVs (\SI{17.44}{m} vs. \SI{10.74}{m}). Over 75\% of the car models has $\delta_0$ above the HDV $P_{85}$.  Notably, the larger $\delta_0$ exists in both minimum and maximum headway (i.e., ACC does not exhibit two distinguished clusters, see Figure \ref{time-gap-jam-spacing_dist}(b)).  

Figure \ref{FD parameters} shows the distribution of capacity, $k_j$, and $w$ compared to the HDV reference.  Clearly, the ACC capacity distribution exhibits two clusters, where the larger (smaller) value group represent the capacity upper (lower) bounds of the ACC systems resulting from the minimum (maximum) headway.  One can see that the ACC systems can reach a very high value. For example, the three highest capacity values are 3593$veh/h$ (H-min), 3430$veh/h$ (X-min), and 3322$veh/h$ (W-min), while the 85 percentile capacity in HDVs is 2581$veh/h$. On the other hand, ACC will result in very small capacity under the maximum headway. For example, it is 1323$veh/h$ for Model S-max, 1380$veh/h$ for C-max, and 1387$veh/h$ for X-max, which is about 15\% lower than HDV average. 

The ACC jam density $k_j$ varies in a much larger range than HDVs with a much smaller lower bound, reaching \SI{50}{veh/km}. Particularly, a large proportion of ACC systems 
have a $k_j$ below the lower bound of the HDVs, 
corresponding to ACC systems with large $\delta_0$ values in Figure \ref{time-gap-jam-spacing_dist}(b). 

The ACC wave speed $w$ varies largely (\SI{-103.8}{km/h} to \SI{-14.1}{km/h}  for minimum headway, and \SI{-47.4}{km/h} to \SI{-8.0}{km/h} for maximum headway, here the two headway settings are distinguished to better show the respective clusters) 
and the value can be much larger than the typical values of HDVs (-20$km/h$ to -10$km/h$ per \citep{chen2012behavioral,laval2010mechanism,duret2011passing,chen2014periodicity}).  In fact, the value of $w$ is mostly larger than 20$km/h$, especially for minimum headway setting.  
Particularly, the fastest wave, $w=-103.8km/h$, is from system G-min, resulting from a small $\tau_0$ (0.61$s$) and large $\delta_0$ (17.44$m$).  Apparently, a fast propagating wave is a distinct feature of ACC. It indicates  that, once a state transition occurs (e.g., a bottleneck is activated),  it can potentially propagate upstream in an extreme fast speed and therefore might have severer impacts (e.g., a human driver may overreact).  Nevertheless, it should be noted that the wave speed here is projected from equilibrium data points, and it represents the \textit{average} wave speed when traffic transitions from one equilibrium state to another. The detailed transition process can be complicated as ACC will be in the non-equilibrium state. This is investigated by the authors in a sequential paper.




\noindent
\begin{minipage}{\linewidth}
\centering
\begin{minipage}[b]{.49\linewidth}
\centering
\footnotesize
\includegraphics[width=.7\textwidth]{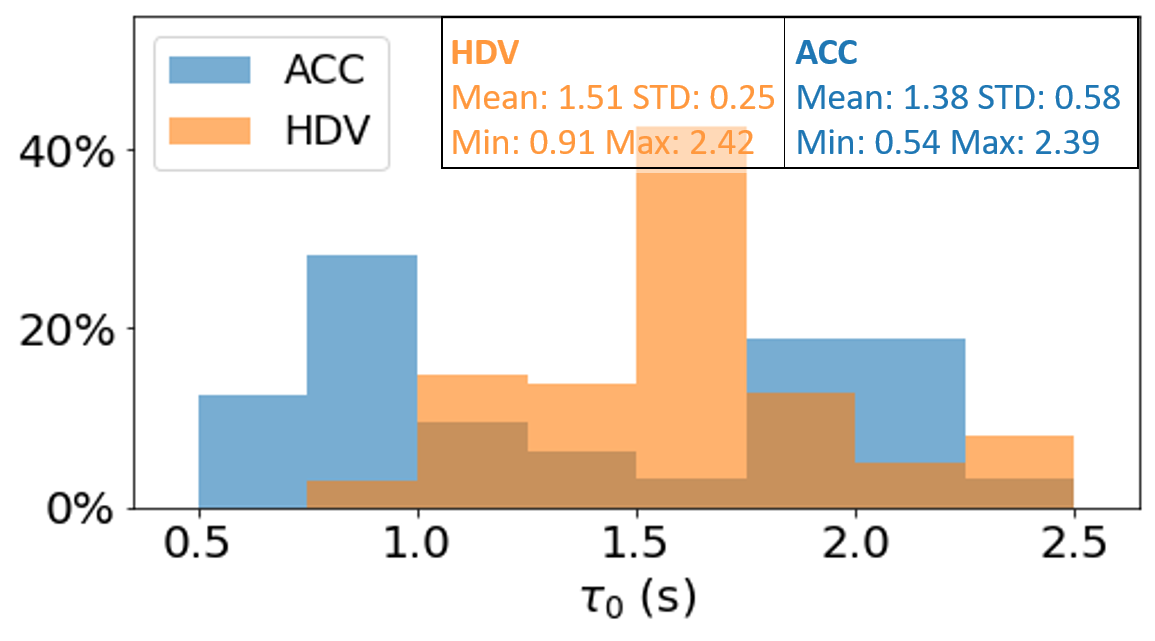}
\\[-6pt]
(a)
\end{minipage}
\begin{minipage}[b]{.49\linewidth}
\centering
\footnotesize
\includegraphics[width=.7\textwidth]{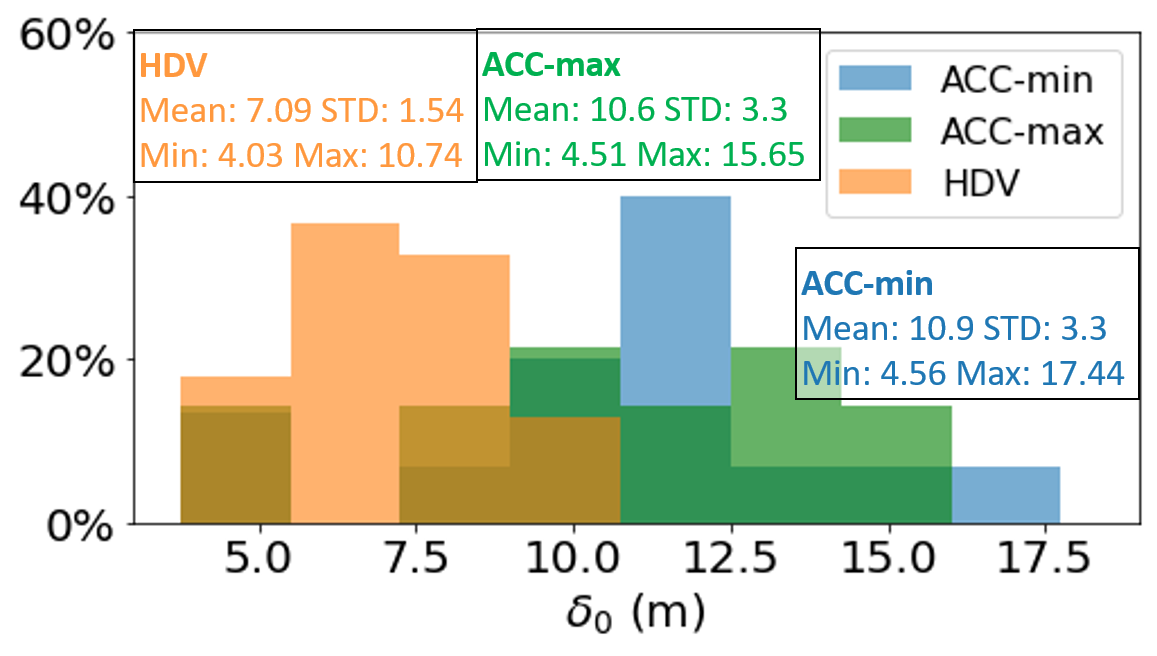}
\\[-6pt]
(b)
\end{minipage} 
\captionof{figure}{$\tau_0$ and $\delta_0$ distribution of the evaluated ACC systems}\label{time-gap-jam-spacing_dist}
\end{minipage}

\noindent
\begin{minipage}{\linewidth}
\centering
\begin{minipage}[b]{.325\linewidth}
\centering
\footnotesize
\includegraphics[width=\textwidth]{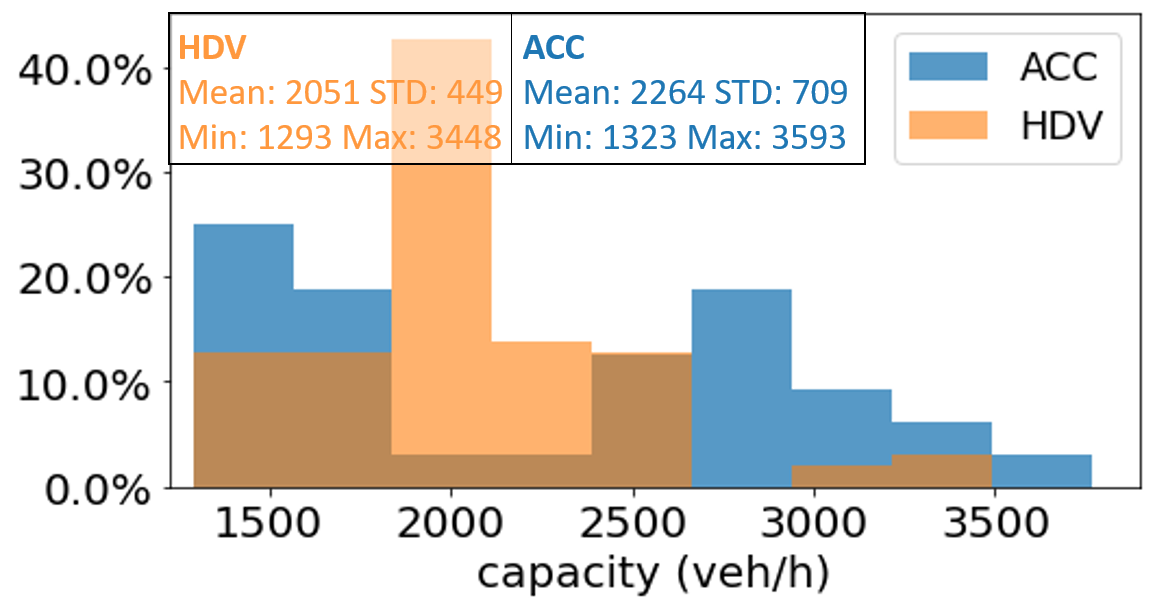}
\\[-2pt]
(a)
\end{minipage}
\begin{minipage}[b]{.325\linewidth}
\centering
\footnotesize
\includegraphics[width=.9\textwidth]{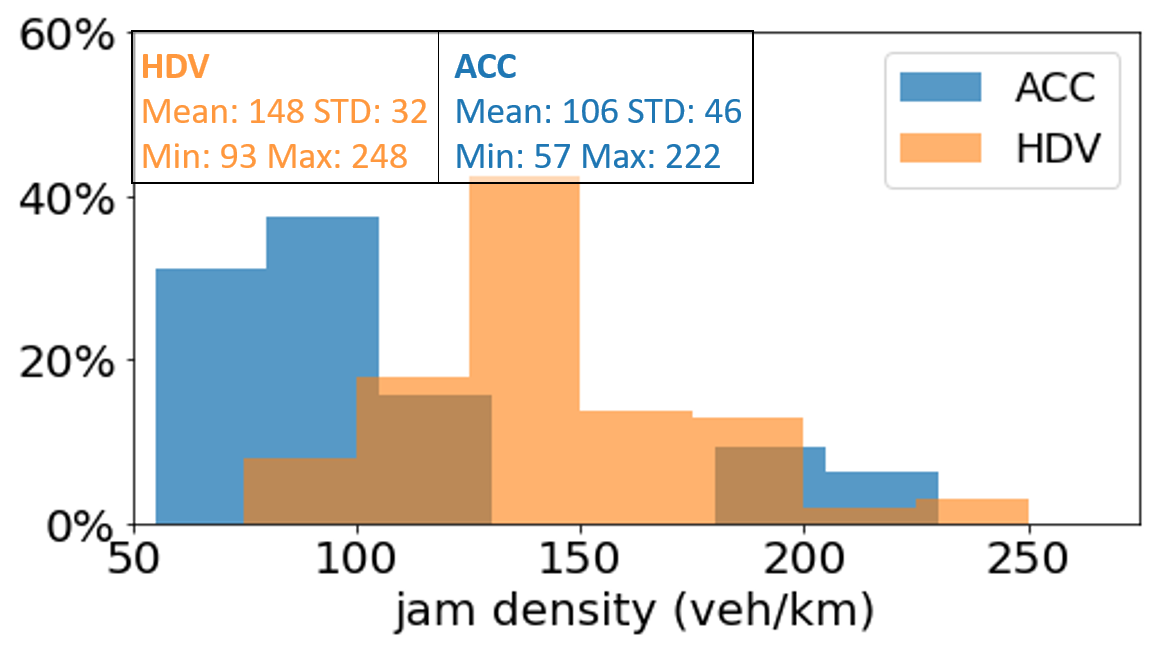}
\\[-2pt]
(b)
\end{minipage} 
\begin{minipage}[b]{.325\linewidth}
\centering
\footnotesize
\includegraphics[width=\textwidth]{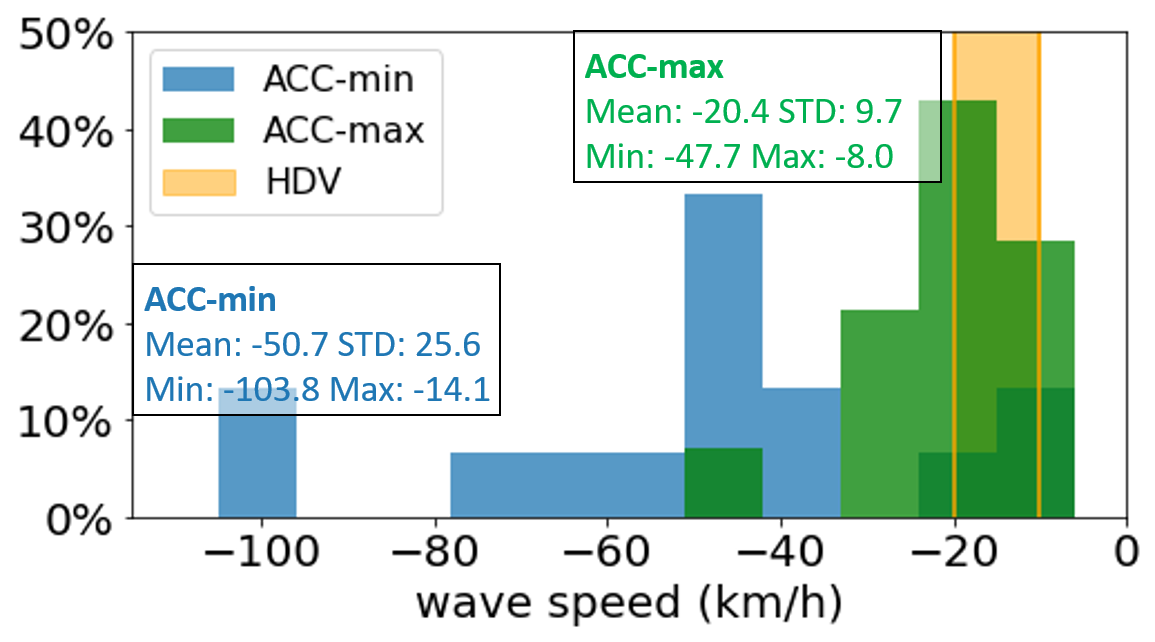}
\\[-2pt]
(c)
\end{minipage} 
\captionof{figure}{capacity, $\delta_0$ and $k_j$ distribution of the evaluated ACC systems}\label{FD parameters}
\end{minipage}

We caution that, for the ACC pool examined, the tested speed ranges often cover medium (or low) to high speed even after consolidation, and the very low speed range was not tested. Therefore, it's unclear whether the estimated $s-v$ relationships still hold outside the tested ranges and whether the projected $\delta_0$ and $k_j$ are realistic.  

Fortunately, the GA experiment has collected some jam spacing data for ACC System X under minmimum headway (measured when the ACC car was at complete stop). Figure \ref{GA_stopping_distance} shows that $\delta_0$ for ACC X-min distributes between \SI{7.7}{m} and \SI{11.56}{m} with the mean at \SI{9.86}{m}.  Interestingly, this is larger than the $\delta_0$ of HDVs; see Figure \ref{NGSIM_qk} for one direct measurement using NGSIM data from \cite{laval2011hysteresis}, which yields a $\delta_0$ about \SI{7}{m}, corresponding to $k_j=$\SI{142}{veh/km}.  Moreover, the measured $\delta_0$ is smaller than the projection from the $s-v$ relationship estimated for low to high speeds (\SI{9.86}{m} vs. \SI{15.00}{m}); see Figure \ref{SystemX_different_sources}.  Note that, for the GA experiment data, if the stopping condition is excluded, it yields a $s-v$ estimation consistent with the MA and OpenACC experiments, although the tested speed ranges differ; see Figure \ref{SystemX_different_sources}(a). However, it differs ($\tau_0$ larger and $\delta_0$ smaller) if the data from stopping condition is included; see Figure \ref{SystemX_different_sources}(b). 

Based on the results above, we conjecture that $s-v$ relationship for system X is not a single linear relationship for the whole speed range. Instead, it may be piece-wise linear or non-linear but linearity is a good approximation in the tested speed range. Moreover, the conjecture likely holds for ACC vehicles in general. The GA measurement of $\delta_0$ suggests that, while the projected $\delta_0$ based on medium(or low) to high speeds is likely larger than the actual jam spacing, it is probably still true that the actual jam spacing values of ACC systems are significantly larger than HDVs. Of course, further research on the very low speed condition (0-\SI{5}{m/s}) is desired to test our hypothesis and complete the $s-v$ relationship.

\noindent
\begin{minipage}{\linewidth}
\centering
\begin{minipage}[b]{.49\linewidth}
\centering
\small
\includegraphics[height=115pt]{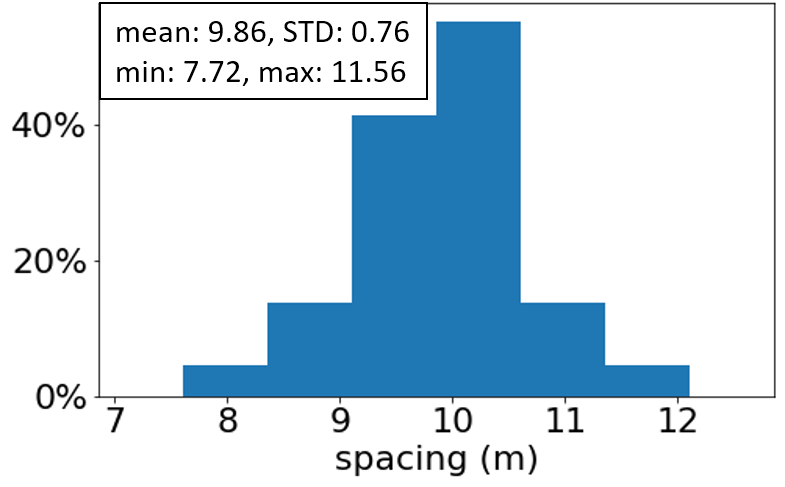}
\end{minipage} 
\captionof{figure}{Spacing distribution of the stopping points in GA experiments}
\label{GA_stopping_distance}
\end{minipage}

\noindent
\begin{minipage}{\linewidth}
\centering
\small
\includegraphics[height=115pt]{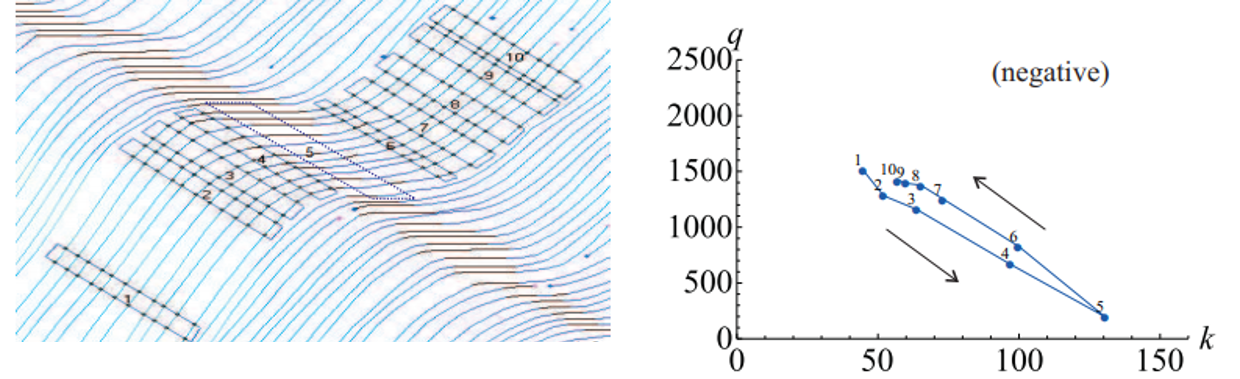}
\captionof{figure}{Direct HDV measurement from NGSIM data \citep{laval2011hysteresis}}
\label{NGSIM_qk}
\end{minipage}

\noindent
\begin{minipage}{\linewidth}
\centering
\begin{minipage}[b]{.49\linewidth}
\centering
\small
\includegraphics[height=150pt]{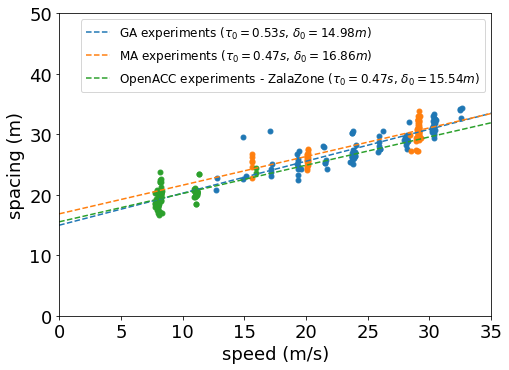}
\\[-6pt]
(a) excluding stopping points
\end{minipage}
\begin{minipage}[b]{.49\linewidth}
\centering
\small
\includegraphics[height=150pt]{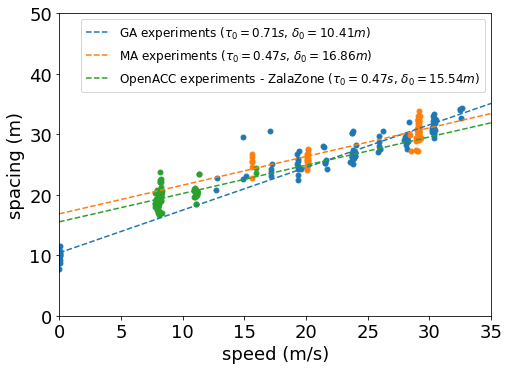}
\\[-6pt]
(b) including stopping points
\end{minipage}
\captionof{figure}{Estimated $s-v$ relationships of the GA data}
\label{SystemX_different_sources}
\end{minipage}

\subsection{\textbf{The general $s-v$ design of ACC}}

We examine the relationship between $\tau_0$ and $\delta_0$.  Figure \ref{tau0_vs_delta0} shows the ($\tau_0$, $\delta_0$) of the ACC pool.  In the minimum headway, $\tau_0$ is usually quite small (mostly below \SI{1}{s}) and the $\delta_0$ is large, resulting in most points in Quadrant II. Interestingly, $\delta_0$ is usually larger if $\tau_0$ is smaller; see the circle markers and the orange dash line (p-value = 0.00).  This is expected as a larger $\delta_0$ can compensate the effect of a smaller $\tau_0$ to increase spacing and thus improve safety, which is particularly critical under the minimum headway setting.  For the maximum headway, the ACC systems are often characterized with large $\tau_0$ and large $\delta_0$; see most points in Quadrant I. Interestingly, some systems have a very large $\delta_0$ even though the $\tau_0$ is already large (see those in the upper right of quadrant I), suggesting that the spacing is profoundly larger than HDVs.  Under the maximum headway, the correlation between  $\tau_0$ and $\delta_0$ is not significant. This is probably because safety is not concerning when the $\tau_0$ are already large. 

Notably, from the perspective of ACC design, it appears reasonable to have a larger $\delta_0$ for small $\tau_0$, as it helps to avoid excessively small spacing in high speed.  Otherwise, it will be too dangerous.  
However, with a very large $\delta_0$, the ACC spacing in very low speed could significantly exceed HDV traffic, which will reduce the storage capacity of roads and may invite surrounding vehicles to cut-in.  Therefore, the result here further supports our conjecture above that the $s-v$ for ACC is not a single linear relationship for the entire speed range.  Our conjecture is also consistent with \cite{shi2021empirical}, which found that the calibrated ACC controller varies with speed.  Further research along this line is needed.

\noindent
\begin{minipage}{\linewidth}
\centering
\includegraphics[width=.5\textwidth]{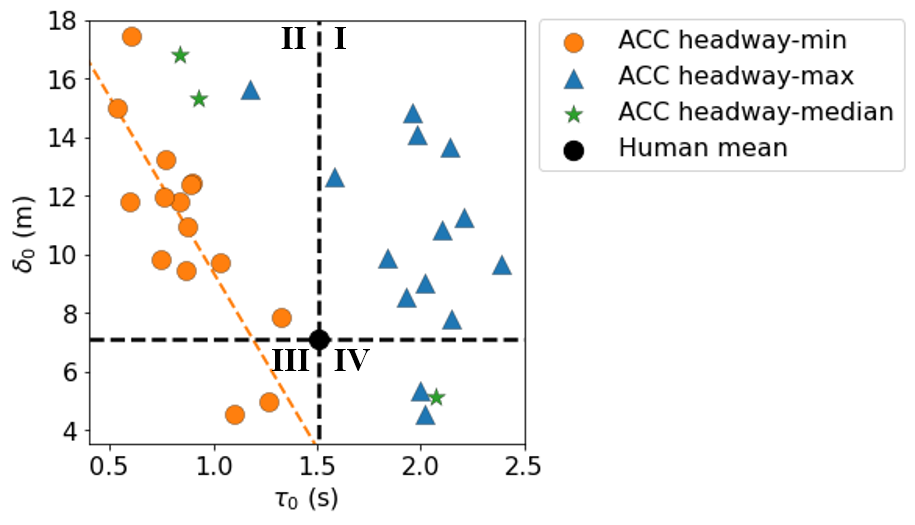}
\captionof{figure}{$\tau_0$ and $\delta_0$ pairs of the evaluated ACC systems}
\label{tau0_vs_delta0}
\end{minipage}
\section{\textbf{Conclusions and discussions}}\label{conclusion} 


In this paper, we have studied the equilibrium behaviors of ACC vehicles based on empirical experiments worldwide and estimated the $s-v$ relationships and FDs for 17 ACC car models from eight manufacturers.  We have proposed a procedure that directly quantifies the ACC behaviors in equilibrium condition and then compared the $s-v$ relationships of the tested ACC systems.  We find that the $s-v$ relationship varies with headway setting but remains unchanged when an ACC vehicle is in different positions of a platoon or under different engine modes.  Across different ACC car models from the same manufacturers, $s-v$ can remain the same or vary. 

Furthermore, we have analyzed the features of the $s-v$ and FDs of the tested ACC pools.  We find that the $s-v$ relationships of ACC systems fit well with a linear formulation in the tested medium (or low) to high speed range (7-\SI{33}{m/s}). However, the $s-v$ likely differs in the very low speed condition ($<$\SI{5}{m/s}). Additionally, the ACC $s-v$ and FDs are very different from HDVs; i.e., at minimum headway, ACC spacing can be much smaller than HDVs and result in much larger flow.  \RoundTwo{The equilibrium capacities in excess of 3500 vehicles per hour are observed. We do note that here the capacity indicates the value in equilibrium status and the actual road capacity observed on the road (flow over a significant period of time) may be different, if the controller is not string stable. This is because for a unstable controller, the traffic condition in real-world traffic can easily transition from equilibrium to non-equilibrium due to the common disturbances and non-equilibrium will dominate in real-world traffic.  } At maximum headway, ACC spacing is often much larger than HDVs and flow is much smaller.  Besides, the direct measurement of $\delta_0$ for ACC system X shows that ACC jam spacing is much smaller than HDVs, which will reduce road storage (jam density is smaller).  Moreover, ACC wave speed is distributed in a very large range, and a significant proportion of ACC systems have extremely large values. \RoundTwo{ The equilibrium wave speed can be as fast as 100 kilometers per hour. A fast back-propagating wave will leave less reaction time for the follower to response to the changing traffic status (e.g., from free flow to congestion), which may impose safety hazard. Note that the equilibrium wave speed will likely differ from the transient wave speed, which captures the propagation of speed change from one vehicle to another. We are currently studying such differences. }

Note that in our analysis, we have observed that spacing variance at a given speed may be significant in some cases, translating to significant flow variation.  Figure \ref{speed_bin}(a) shows an example of the observed $s-v$ relationship on Tesla Model S with maximum headway and Figure \ref{speed_bin}(b) shows the spacing distribution of one speed bin.  Measurement error can play a role on that.  We find that the spacing variation within a speed bin decreases with sample size and converges to a reasonable level (around \SI{1.5}{m}); see Figure \ref{speed_bin_STD}. The OpenACC-AstaZero data (red points) has a consistently lower variance than the other data sources.  This is likely because AstaZero has a measurement system with larger accuracy (mean location error was claimed to be \SI{0.02}{m} per \cite{makridis2021openacc}, compared to the error mean around  \SI{1.36}{m} to \SI{1.68}{m} for the MA/Work/GA/OpenACC-ZalaZon experiments which used similar GPS devices). Note that spacing variation is still observed even we impose stricter criteria on the equilibrium conditions 
and sample size from the highly accurate AstaZero data.  Two factors can contribute to that.  Firstly, in real-world traffic, due to the changing environment (e.g., aerodynamic, gradient, pavement condition), it is extremely difficult, if possible at all, for a vehicle to stay in the absolute equilibrium condition for a significant period of time.  Thus, the equilibrium intervals measured will almost inevitably have some variation in the traffic states (stricter criteria will help to reduce the variation to a certain extent but couldn't fully eliminate that).  Additionally, the ACC systems rely on sensors (mostly radar and/or camera) to provide the distance and speed measurement of the leader, which have measurement errors.  Such errors could result in mis-perception in ACC 
and thus contribute to the spacing variation. Besides the measurement errors, it's also possible that ACC design embeds a buffer in the $s-v$ so that it will not be overly sensitive to disturbances. Future research along this line, particularly on the low-level execution of ACC, is desired to investigate this issue.    

\RoundTwo{Future research is needed to test the ACC systems in the very low speed conditions (0-\SI{5}{m/s}) in order to complete the estimation of $s-v$ and FD. It's also desirable to conduct the field test at more speed levels so that one has more than three bins to estimate the $s-v$ relationship, which will provide more accurate results.  Besides, field tests are desired to investigate the safety impacts of the extremely large wave speed on vehicles following ACC, especially human drivers.}  

Another line of future research is to extrapolate the impacts of the general ACC. This study has shown that commercial ACC systems are very different from each other and from HDVs, which will produce  profound impacts on traffic flow.  Future research is needed to investigate the causes of such differences to enable the extrapolation.  We conjecture that some vehicle features can play a significant role in the differences, such as \RoundTwo{the low-level controller \cite{zhou2021impact} } , power to weight ratio, and the sensing type (e.g., radar vs. camera vs. Lidar). This issue is under investigation of the research team.  Besides, the extrapolation also requires research on the interactions between ACC and HDVs.

\noindent
\begin{minipage}{\linewidth}
\centering
\begin{minipage}[b]{.49\linewidth}
\centering
\footnotesize
\includegraphics[width=.75\textwidth]{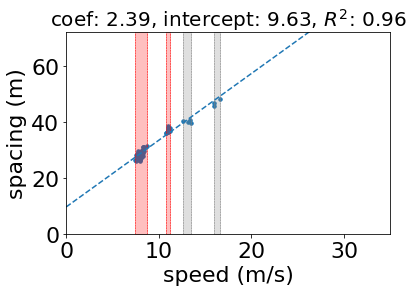}
\\[-6pt]
(a) $s-v$ relationship
\end{minipage}
\begin{minipage}[b]{.49\linewidth}
\footnotesize
\centering
\includegraphics[width=.75\textwidth]{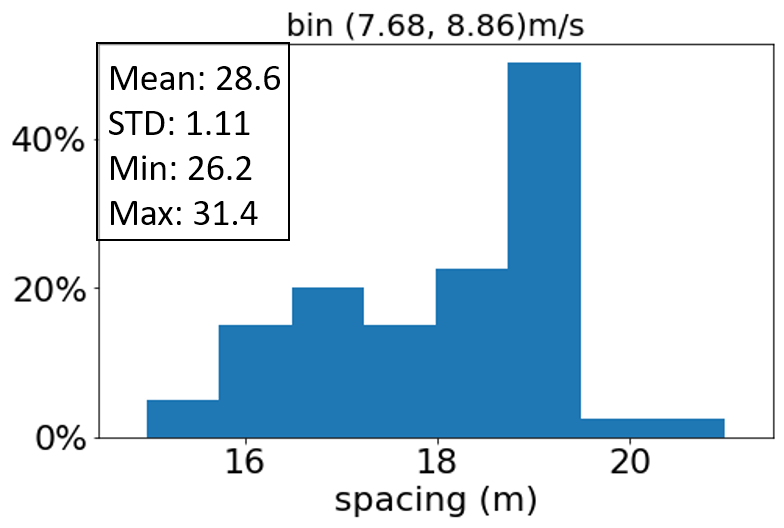}
\\[-6pt]
(b) distribution of spacing between \SI{7.68}{m/s} to \SI{8.86}{m/s}
\end{minipage} 
\captionof{figure}{An example of speed bin recognition and spacing distribution in the bins}
\label{speed_bin}
\end{minipage}

\noindent
\begin{minipage}{\linewidth}
\centering
\begin{minipage}[b]{.49\linewidth}
\footnotesize
\centering
\includegraphics[width=.75\textwidth]{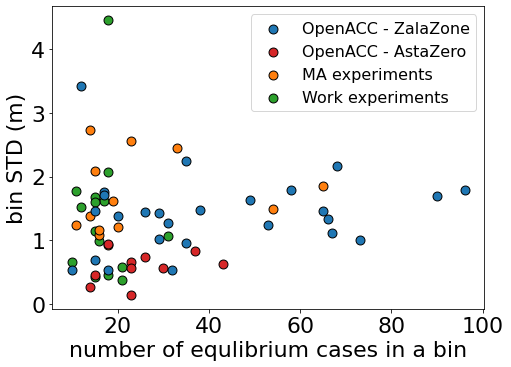}
\\[-2pt]
\end{minipage} 
\captionof{figure}{The relationship of the bin standard deviation with sample size and speed}
\label{speed_bin_STD}
\end{minipage}

\section*{\textbf{\uppercase{Acknowledgement}}}
This research was sponsored by the National Science Foundation through Awards CMMI 1826162, CMMI 1932921, CMMI 1826003 and CPS 1932451. 
The authors would like to gratefully thank George Gunter, Dr. Dan Work, Dr. Michail Makridis, and Dr. Biagio Ciuffo for sharing the treasured experiment data with the community and discussing the information of data collection in their experiments with the team.  


\bibliographystyle{elsarticle-harv}\biboptions{authoryear}
\bibliography{references}

\newpage 
\appendix

\newpage
\section{Table of comparing the same ACC systems}
\setcounter{table}{0}

\begin{table}[!h]
\caption{ACC Vehicles in Different Position of A Platoon (based on MA experiment)}\label{compare_same_system}
\centering
\footnotesize{
\begin{tabular}{|>{\centering\arraybackslash}m{.8cm}|>{\centering\arraybackslash}m{1.1cm}|>{\centering\arraybackslash}m{.9cm}|>{\centering\arraybackslash}m{.8cm}|>{\centering\arraybackslash}m{.95cm}|>{\centering\arraybackslash}m{.9cm}|>{\centering\arraybackslash}m{.8cm}|>{\centering\arraybackslash}m{.95cm}|>{\centering\arraybackslash}m{1cm}|>{\centering\arraybackslash}m{1.75cm}|>{\centering\arraybackslash}m{1.75cm}|}
\hline
\multirow{2}{*}{system} & \multirow{2}{*}{headway} &\multirow{2}{*}{mode} & \multicolumn{3}{c|}{ACC1}   & \multicolumn{3}{c|}{ACC2}   & \multirow{2}{1.9cm}{\centering $\tau_0$ difference (p-value)} & \multirow{2}{1.9cm}{\centering $\delta_0$ difference (p-value)} \\ \cline{4-9}
                   &     &                       & $\tau_0$ (s) & $\delta_0$ (m) & sample  & $\tau_0$ (s) & $\delta_0$ (m) & sample  &                                    &                                     \\ \hline
W   &      min            & normal                & 0.62 & 12.21  & 50          & 0.52 & 16.03  & 42          & -0.10 (0.21)                        & 3.82 (0.06)                         \\ \hline
W     &     3           & normal                & 0.92 & 14.55   & 31          & 0.98 & 15.48  & 25          & 0.06 (0.67)                       & 0.94 (0.79)                         \\ \hline

X    &       min          & normal                & 0.46 & 17.32  & 41          & 0.49 & 16.16  & 37          & 0.03 (0.68)                       & -1.16 (0.52)                         \\ \hline
X    &        3         & normal                & 0.83 & 16.80  & 18          & 0.95 & 15.11  & 9          & 0.12 (0.31)                        & -1.70 (0.49)                        \\ \hline

Y   &       min           & normal                & 1.09 & 4.71   & 13           & 1.20 & 4.32   & 4           & 0.10 (0.38)                        & -0.39 (0.89)                        \\ \hline
Y     &         3       & normal                & 2.15 & 3.42   & 15          & 2.06 & 6.53   & 6           & -0.09 (0.57)                       & 3.11 (0.39)                         \\ \hline

Y   &       min           & sport                 & 1.06 & 4.70   & 9          & 1.14 & 3.04   & 4          & 0.08 (0.31)                       & -1.66 (0.40)                         \\ \hline
Y    &      3           & sport                 & 2.04 & 5.39   & 21          & 2.10  & 4.40   & 3          & -0.06 (0.74)                        & -0.99 (0.85)                        \\ \hline

Z   &       min           & normal                & 1.31 & 1.62   & 5          & 0.64 & 17.71  & 3          & -0.67 (0.26)                        & 16.09 (0.24)                         \\ \hline
Z  &        max           & normal                & 2.34 & 12.80  & 4          & 2.10 & 14.87  & 4           & -0.24 (0.68)                       & 2.08 (0.87)                         \\ \hline

Z  &    min              & power                 & 0.75 & 14.01  & 12          & 1.09 & 8.53  & 8          & 0.33 (0.20)                       & -5.49 (0.30)                         \\ \hline
Z  &    max              & power                 & 1.97 & 18.12  & 13          & 1.85 & 20.81  & 7           & -0.12 (0.75)                       & 2.68 (0.74)                        \\ \hline

\end{tabular}
}
\end{table}

\begin{table}[H]
\caption{ACC under different engine modes (based on MA experiment)}\label{compare_engine}
\centering
\begin{threeparttable}
\footnotesize{
\begin{tabular}{|>{\centering\arraybackslash}m{.9cm}|>{\centering\arraybackslash}m{1.2cm}|>{\centering\arraybackslash}m{1cm}|>{\centering\arraybackslash}m{1cm}|>{\centering\arraybackslash}m{1.1cm}|>{\centering\arraybackslash}m{1cm}|>{\centering\arraybackslash}m{1cm}|>{\centering\arraybackslash}m{1.1cm}|c|c|}
\hline
\multirow{2}{*}{System} & \multirow{2}{*}{Headway} & \multicolumn{3}{c|}{Normal}   & \multicolumn{3}{c|}{Sport/Power}   & \multirow{2}{1.9cm}{\centering $\tau_0$ difference (p-value)} & \multirow{2}{1.9cm}{\centering $\delta_0$ difference (p-value)} \\ \cline{3-8}
                        &                      & $\tau_0$ (s) & $\delta_0$ (m) & sample  & $\tau_0$ (s) & $\delta_0$ (m) & sample  &                                     &                                     \\ \hline
Y & min & 1.12 & 4.53  & 17 & 1.10 & 3.83  & 13  & -0.02 (0.77)  & -0.70 (0.70) \\ \hline
Y & 3 & 2.13 & 4.26  & 20 & 2.05 & 5.21  & 24  & -0.08 (0.27)  & 0.95 (0.65)  \\ \hline
Z & min & 1.03 & 8.25 & 8 & 0.84  & 12.60 & 20 & -0.19 (0.51) & 4.35 (0.49)  \\ \hline
Z & max & 2.27 & 12.79 & 8 & 1.93 & 18.93 & 20 & -0.34 (0.38) & 6.14 (0.48)  \\ \hline
\end{tabular}
}
\end{threeparttable}
\end{table}

\newpage
\section{Estimated $s-v$ relationships of the tested ACC pool}
\setcounter{table}{0}
\setcounter{figure}{0}

\begin{table}[h]
\caption{Spacing-speed regression results of the evaluated ACC systems}\label{full_sv_regression_table}
\centering
\footnotesize{
\begin{tabular}{|>{\centering\arraybackslash}m{2cm}|>{\centering\arraybackslash}m{2cm}|>{\centering\arraybackslash}m{2cm}|>{\centering\arraybackslash}m{2cm}|>{\centering\arraybackslash}m{2cm}|>{\centering\arraybackslash}m{2cm}|}
\hline
system & headway & $\tau_0$ (s) & $\delta_0$ (m) & $R^2$   & sample size \\ \hline
A & Min & 0.89 & 12.39 & 0.91 & 68  \\ \hline
A & Max & 1.96 & 14.83  & 0.98 & 43  \\ \hline

B & Min & 0.87 & 10.92  & 0.70 & 59  \\ \hline
B & Max & 2.10 & 10.83 & 0.84 & 47  \\ \hline

C & Min & 0.90 & 12.45 & 0.78 & 73  \\ \hline
C & Max & 2.01 & 16.17 & 0.89 & 78  \\ \hline

D & Min & 0.84 & 11.77 & 0.95 & 84  \\ \hline
D & Max & 1.98 & 14.09 & 0.85 & 52  \\ \hline

E & Min & 1.27 & 4.97  & 0.95 & 54  \\ \hline
E & Max & 2.02 & 9.00  & 0.96 & 52  \\ \hline

F & Min & 0.77 & 13.24 & 0.93 & 57  \\ \hline
F & Max & 2.02 & 4.51  & 0.96 & 55  \\ \hline

G & Min & 0.61 & 17.44 & 0.86 & 54  \\ \hline
G & Max & 2.00 & 5.36  & 0.97 & 51  \\ \hline

H & Min & 0.60 & 11.78 & 0.95 & 63  \\ \hline
H & Max & 1.84 & 9.85  & 0.99 & 46  \\ \hline

W    & Min & 0.75 & 9.79 & 0.90 & 137        \\\hline
W    & Median & 0.93 & 15.33 & 0.73 & 56       \\\hline

W    & Max & 1.18 & 15.65 & 0.63 & 99        \\\hline

X   & Min & 0.54 & 15.00 & 0.91 & 269       \\\hline
X    & Median & 0.84 & 16.78 & 0.92 & 27       \\\hline

X    & Max & 2.21 & 11.27 & 0.83 & 119       \\\hline

Y & Min & 1.10 & 4.56  & 0.96 & 30  \\ \hline
Y & Median & 2.15 & 7.78  & 0.98 & 45 \\ \hline

Model S   & Min & 1.03 & 9.72 & 0.77 & 81       \\\hline
Model S   & Max & 2.39 & 9.63 & 0.96 & 89       \\\hline

A6    & Min & 0.76 & 11.96 & 0.85 & 77          \\\hline

X5      & Min & 0.87 & 9.42 & 0.93 & 102         \\\hline

I3 s    & Min & 1.33 & 7.86 & 0.47 & 100           \\\hline
I3 s    & Max & 2.15 & 7.78 & 0.60 & 113           \\\hline

GLE450   & Max & 1.93 & 8.56 & 0.73 & 35            \\\hline

I-Pace   & Max & 1.57 & 12.67 & 0.86 & 52             \\\hline

\end{tabular}
}
\end{table}

\noindent
\begin{minipage}{\linewidth}
\centering
\begin{minipage}[b]{.325\linewidth}
\centering
\footnotesize
\includegraphics[width=\textwidth]{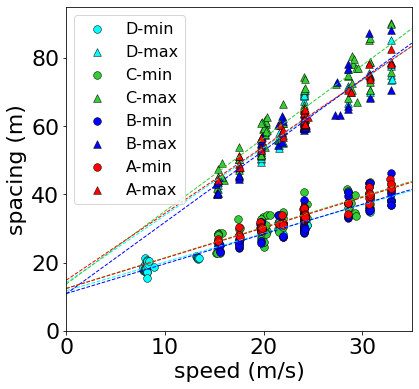}
\\[-2pt]
(a) Manufacturer 1
\end{minipage} 
\begin{minipage}[b]{.325\linewidth}
\centering
\footnotesize
\includegraphics[width=\textwidth]{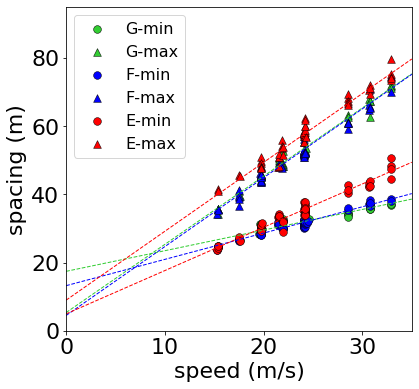}
\\[-2pt]
(b) Manufacturer 2
\end{minipage}
\begin{minipage}[b]{.325\linewidth}
\centering
\footnotesize
\includegraphics[width=\textwidth]{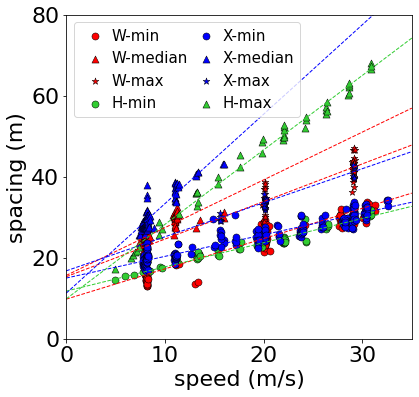}
\\[-2pt]
(c) Manufacturer 3
\end{minipage}


\captionof{figure}{$s-v$ relationships for Manufacturers 1-3 from the consolidated data}\label{sv_Manufacturer}
\end{minipage}

\noindent
\begin{minipage}{\linewidth}
\centering
\begin{minipage}[b]{.325\linewidth}
\centering
\footnotesize
\includegraphics[width=\textwidth]{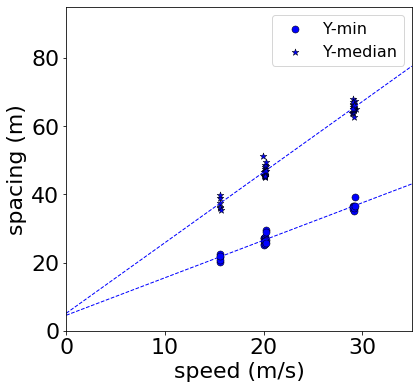}
\\[-2pt]
(a) Manufacture 4
\end{minipage} 
\begin{minipage}[b]{.325\linewidth}
\centering
\footnotesize
\includegraphics[width=\textwidth]{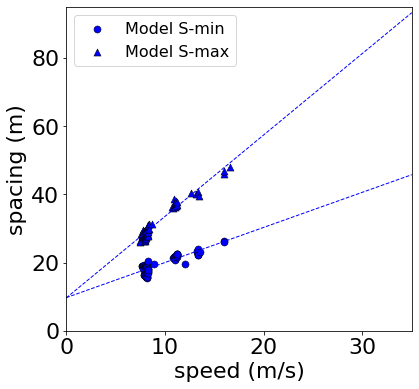}
\\[-2pt]
(b) Tesla
\end{minipage} 
\begin{minipage}[b]{.325\linewidth}
\centering
\footnotesize
\includegraphics[width=\textwidth]{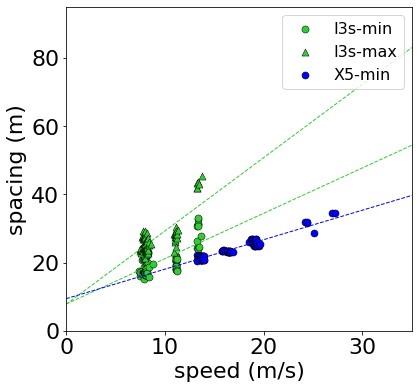}
\\[-2pt]
(c) BMW
\end{minipage} 

\begin{minipage}[b]{.325\linewidth}
\centering
\footnotesize
\includegraphics[width=\textwidth]{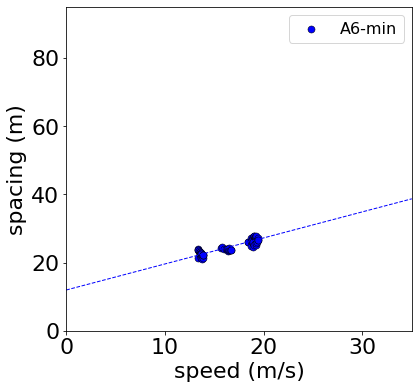}
\\[-2pt]
(d) Audi
\end{minipage} 
\begin{minipage}[b]{.325\linewidth}
\centering
\footnotesize
\includegraphics[width=\textwidth]{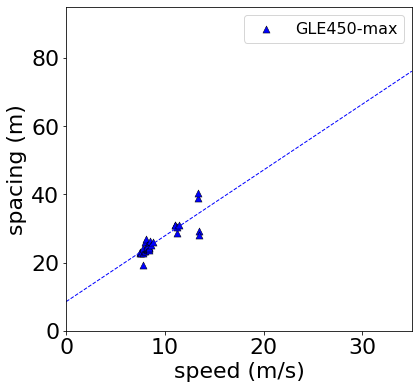}
\\[-2pt]
(e) Mercedes
\end{minipage} 
\begin{minipage}[b]{.325\linewidth}
\centering
\footnotesize
\includegraphics[width=\textwidth]{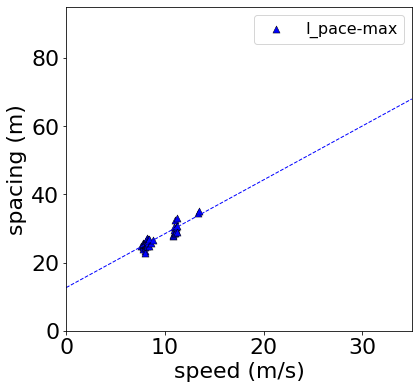}
\\[-2pt]
(f) Jaguar
\end{minipage} 
\captionof{figure}{$s-v$ relationships from the consolidated data (other manufacturers) }\label{sv_Manufacturer_other}
\end{minipage}

\end{document}